\documentclass[11pt]{article}
\pdfoutput=1

\usepackage{a4wide}
\usepackage[fleqn]{amsmath}

\RequirePackage{amsmath,amssymb,amsxtra,amsthm}

\RequirePackage[T1]{fontenc}
\RequirePackage[utf8]{inputenc}

\RequirePackage{mathrsfs}
\RequirePackage{mathpazo}

\RequirePackage{setspace}
\RequirePackage{array}
\RequirePackage{booktabs}

\RequirePackage{braket}

\RequirePackage[pdftex,final]{graphicx}
\graphicspath{{Plots/}}

\usepackage{cite}

\newcommand{\be}{\begin{equation}}
\newcommand{\ee}{\end{equation}}
\newcommand{\bea}{\begin{eqnarray}}
\newcommand{\eea}{\end{eqnarray}}
\newcommand{\IR}{\mathbb{R}}
\newcommand{\IC}{\mathbb{C}}
\newcommand{\IZ}{\mathbb{Z}}

\newcommand{\IN}{\mathbb{N}}

\newcommand{\cF}{\mathcal{F}}

\newcommand{\cS}{\mathcal{S}}

\newcommand{\cH}{\mathcal{H}}

\newcommand{\cM}{\mathcal{M}}

\newcommand{\cO}{\mathcal{O}}
\newcommand{\cR}{\mathcal{R}}
\newcommand{\cT}{\mathcal{T}}

\newcommand{\de}{\mathrm{d}}

\newcommand{\I}{\mathrm{i}}
\newcommand{\zetastar}{\zeta^\star}
\newcommand{\Rstar}{\mathcal{R}^\star}

\numberwithin{equation}{section}
\numberwithin{table}{section}
\numberwithin{figure}{section}

\author{
  \begin{minipage}{0.97\linewidth}
    \vspace{1cm}
    \begin{center}
      \begin{small}
        \textbf{Carlo Angelantonj}$^{1,3}$, \textbf{Ioannis Florakis}$^{2}$ and  \textbf{Boris Pioline}$^{3,4}$
     \end{small}
    \end{center}
    \vspace{.3cm} \hspace{1.3cm}\begin{minipage}{.75\linewidth}
      {\it \begin{footnotesize}
          \begin{itemize}
          \item[${}^1$] Dipartimento di Fisica Teorica, Universit\`a di Torino, and INFN Sezione di Torino
          \\
            Via P. Giuria 1, 10125 Torino, Italy
          \item[${}^2$] Max-Planck-Institut f\"{u}r Physik,\\
	    Werner-Heisenberg-Institut,
          80805 M\"{u}nchen, Germany
          \item[${}^3$] CERN Dep PH-TH, 1211 Geneva 23, Switzerland
         \item[${}^4$] Laboratoire de Physique Th\'eorique et Hautes Energies, CNRS UMR 7589,
         \\
         Universit\'e Pierre et Marie Curie - Paris 6, 4 place Jussieu,
         75252 Paris cedex 05, France
          \end{itemize}
        \end{footnotesize}}
    \end{minipage}
    \vspace{1cm}
  \end{minipage}
}

\date{}

\title{\vspace{3cm}
  \begin{huge}
    \textbf{A new look at one-loop integrals \\[2mm] in string theory}
  \end{huge}
}

\begin{document}

\begin{titlepage}
  \maketitle
  \thispagestyle{empty}

  \vspace{-14cm}
  \begin{flushright}
    CERN-PH-TH/2011-259\\  
    DFTT 23/2011\\
    MPP-2011-11911\\
    arXiv:1110.5318v2
   \end{flushright}

  \vspace{11cm}

  \begin{center}
    \textsc{Abstract}\\
  \end{center}

We revisit the evaluation of one-loop modular integrals in string theory, employing new 
methods that, unlike the traditional 'orbit method',  keep  T-duality manifest throughout. In particular, we apply the Rankin-Selberg-Zagier approach to cases where the integrand function grows at most polynomially in the IR. Furthermore, we introduce new techniques in the case where `unphysical tachyons'  contribute to the one-loop couplings. These methods can  be viewed as a modular invariant version of dimensional regularisation. As an example, we treat one-loop BPS-saturated couplings involving the $d$-dimensional Narain lattice and the invariant Klein $j$-function, and relate them to (shifted) constrained Epstein Zeta series of ${\rm O} (d,d; \mathbb{Z})$.
In particular,  we recover the well-known results for $d=2$  in a few easy steps.

\vfill
{\small
\begin{itemize}
\item[E-mail:] {\tt carlo.angelantonj@unito.it}\\ {\tt florakis@mppmu.mpg.de}\\
{\tt pioline@lpthe.jussieu.fr}
\end{itemize}
}
\vfill

\end{titlepage}

\setstretch{1.1}

\tableofcontents


\section{Introduction}

In essence, closed string theory is a quantum field theory of infinitely many fields, obtained by tensoring two infinite towers of left-moving and right-moving excitations subject to a level-matching constraint \cite{Schwarz:1982jn}. As a result, any closed scattering amplitude at one-loop is an integral over two parameters, the Lagrange multiplier $\tau_1\in[-\tfrac12,\tfrac12]$ for the level-matching constraint  and the Schwinger time $\tau_2>0$ parameterising the loop. Due to diffeomorphism invariance on the string world-sheet, the identification of the proper time is not unique and the integrand $F(\tau)$ is a modular function of the complex parameter $\tau\equiv \tau_1+\I\tau_2$ (in general not holomorphic). To avoid an infinite over-counting, the domain of integration is restricted to a fundamental domain $\cF$ for the modular group $\varGamma ={\rm SL}(2,\IZ)$, thereby removing the ultraviolet divergences from the region 
$\tau_2\to 0$ which usually arise in quantum field theory. The usual field theoretical
infrared divergences  from the region
$\tau_2\to\infty$ are in general still present, due the existence of massless particles in the spectrum\footnote{In the following, we restrict our attention to closed string theories without physical tachyons, but allow for `unphysical tachyons', i.e. 
relevant operators which do not satisfy the level-matching condition, such as those present in heterotic models.}. 

In general, computing such `one-loop modular integrals' is a daunting task, (in part) due to the unwieldy shape of the domain $\cF$. For specific amplitudes describing BPS-saturated couplings in the low-energy effective action however, the integrand simplifies and the modular integral can be computed explicitly. One of the simplest  instances occurs for  certain anomaly-related couplings in the ten-dimensional heterotic string theory \cite{Lerche:1987qk}. In this case the integrand is the elliptic genus $F=\varPhi(\tau)$, a weak holomorphic modular form with a singularity at the boundary of $\cF$, and the modular integral can be evaluated by applying Stokes' theorem \cite{Lerche:1987qk,0919.11036}. 

In lower dimensions, however, the low-energy couplings depend non-trivially on the geometric 
moduli of an internal $d$-dimensional torus $T^d$. Indeed the integrand function is 
typically of the 
form $\varPhi(\tau) \varGamma_{(d+k,d)}$, where $\varPhi(\tau)$ is again a weak holomorphic 
modular form and $\varGamma_{(d+k,d)}$ is the partition function of the Narain lattice associated 
to the torus compactification. The integrand is not holomorphic, so the  integral cannot be
reduced to a line integral on the boundary of $\cF$ via Stokes' theorem. Rather, the main 
technique for dealing with such modular integrals in the physics literature has been the 
 `unfolding trick' or `orbit method',  pioneered in \cite{McClain:1986id, O'Brien:1987pn} and generalised in many subsequent works  \cite{0919.11036,
 Dixon:1990pc, Mayr:1993mq, Harvey:1995fq, Bachas:1997mc, stst1, stst2, Kiritsis:1997hf, Kiritsis:1997em, Marino:1998pg}. In a nutshell, this method consists in 
extending the integration domain $ \cF$ to a simpler region (the strip $\tau_1\in[-1/2,1/2], \tau_2>0$, or the full upper half plane $\tau_1\in\IR, \tau_2>0$) at the cost of restricting the sum over momenta and windings to suitable orbits of lattice vectors.  While leading to a very useful expansion at large volume, 
this method has the drawback of obscuring the invariance of the resulting low-energy coupling 
under the automorphism group ${\rm O} (d+k,d;\IZ)$ of the Narain lattice. Although in some simple 
cases it is still possible to rewrite the result in terms of known automorphic forms of 
 ${\rm O} (d+k,d;\IZ)$, this is  in general not easy to achieve. 
For $k=0, \Phi=1$, it was conjectured in \cite{Obers:1999um} that the result of the 
one-loop modular integral could in fact be expressed (in several different ways) 
as a  constrained Epstein zeta series, manifestly invariant under ${\rm O} (d,d;\IZ)$.

The purpose of this note is to introduce a procedure for evaluating modular integrals which keeps manifest any (additional) symmetry of the integrand function. To this end we apply  the Rankin-Selberg-Zagier (RSZ) method, a close relative of the `unfolding method'  which has been 
a standard tool in the mathematics literature (see e.g. \cite{MR993311} for a survey). 
In short, the main idea is to insert a non-holomorphic Eisenstein series in  the 
modular integral of interest, 
\footnote{Throughout the paper, $\de\mu = \tau_2^{-2}\, \de\tau_1 \, \de\tau_2$ denotes the ${\rm SL} (2;\mathbb{Z})$ invariant measure on the hyperbolic plane, normalised so that $\int_\mathcal{F} \de\mu = \pi /3$.}
\be
\label{sketch}
\int_{\cF}  \de\mu\, F(\tau)\quad \longrightarrow \quad
\int_{\cF} \de \mu\,  \sum_{\substack{(m,n)\in \IZ^2,\\ {\rm gcd} (m,n)=1}}  \left( \frac{\tau_2}{|m-n \tau|^2} \right)^{s}   \, F(\tau)\ ,
\ee
compute the latter `deformed' integral by applying the `unfolding trick' to the sum over $m,n$
for large $\Re(s)$,  and finally obtain the desired integral  by analytically continuing the 
result to $s=0$ (where the Eisenstein series is known to reduce to a constant).  
Due to the growth of $F(\tau)$ as $\tau\to\I \infty$ in cases of physical
interest, the integral in \eqref{sketch} is infrared divergent, while the original
Rankin-Selberg method was restricted to functions $F(\tau)$ of rapid
decay at the cusp. Fortunately, the Rankin-Selberg method was extended 
by Zagier in \cite{MR656029} to allow for functions of moderate
growth at the cusp, by introducing a hard infrared cut-off $\tau_2<\cT$ on
the Schwinger time,  unfolding the sum over $m,n$ at finite $\cT$, and 
giving a prescription for renormalising the integral as the infrared cut-off is removed. 
With this renormalisation prescription, one can view the replacement \eqref{sketch}
as a stringy analogue of dimensional regularisation, which preserves modular 
invariance\footnote{Other modular invariant infrared regulators have been proposed, e.g. 
\cite{Kiritsis:1994ta}. It is also possible to regulate the integral by subtracting the non-decaying
part of $F(\tau)$, as in  \cite{Dixon:1990pc}. As we shall see, it is straightforward to relate
these different regularisation schemes. For an early use of hard infrared 
cut-off regularisation methods in string theory, see e.g. \cite{Green:1999pv}.}. 
Moreover, the integral \eqref{sketch} for $s\neq 0$ 
literally arises for certain BPS couplings, {\em e.g.} the $D^4 R^4$ couplings studied in \cite{Green:2010wi}. 
Using the RSZ method, we shall evaluate the renormalised modular integral 
$\int_\cF\, \de\mu \varGamma_{(d,d)}$ exactly for any value of $d$, in terms
of a constrained Epstein zeta series of ${\rm O}(d,d,\IZ)$, thereby proving
the conjecture in  \cite{Obers:1999um}. We shall also recover the celebrated result 
of \cite{Dixon:1990pc} for $d=2$ in just in a few easy steps, illustrating
 the power of this approach.

While the Rankin-Selberg-Zagier method outlined above is very efficient for 
functions $F(\tau)$ of polynomial growth, which is typically the case for BPS
couplings in type II string theory,  it is unfortunately inadequate for functions 
$F(\tau)$ of exponential growth, which typically arise in heterotic amplitudes, 
due to the ubiquitous unphysical tachyon.  More specifically, we are interested
in modular integrals of the form 
\be
\label{intintro2}
\int_\mathcal{F} \de\mu\, \varPhi (\tau ) \, \varGamma_{(d+k,d)} \,,
\ee
where $\varPhi (\tau )$ is a weak holomorphic modular form of weight $w=-k/2$ with 
an essential singularity at the cusp, $\varPhi(\tau)\sim e^{-2\pi\I\kappa\tau} + \cO(1)$
with $\kappa>0$ (for heterotic strings, $\kappa=1$ but our method works equally well for any $\kappa>0$). Since the $\tau_1$-average of the polar part of 
$F(\tau) = \varPhi (\tau ) \, \varGamma_{(d+k,d)}$ vanishes, the divergence of the integral
\eqref{intintro2} is not worse for $\kappa>0$ than for $\kappa=0$, but the 
RSZ method nevertheless fails and one must resort to
different techniques. 

In the second part of this work we shall  develop a new procedure for
dealing with the above class of modular integrals, which relies on representing the 
holomorphic part $\varPhi (\tau )$ of the integrand in terms of a Poincar\'e 
series\footnote{Here, $\varGamma_\infty \subset \varGamma$ is the stabiliser of the cusp 
$\I\infty$, generated by the triangular matrices $\gamma = \begin{pmatrix}1 & n\\ 0 & 1\end{pmatrix}$, $n\in \mathbb{Z}$.}
\be 
\label{poincaintro}
 \varPhi (\tau ) = \sum_{\gamma \in \varGamma_\infty \backslash \varGamma} \psi (\gamma \cdot \tau )\,,
\ee
for a suitable function $\psi (\tau )$, and then applying the unfolding trick to the sum in \eqref{poincaintro}. Our procedure is close in spirit to the Rankin-Selberg method,
in particular it keeps manifest the ${\rm O} (d+k,d;\mathbb{Z})$ symmetry of the Narain lattice,
however note that it  does not involve any auxiliary modular function  for unfolding the fundamental domain as in (\ref{sketch}), rather it uses the function $F$ itself or part of it. 
 
A naive implementation of this idea, however, is hampered by the fact that the Poincar\'e series of 
a modular form of non-positive weight is not absolutely convergent and thus its unfolding is not justified. We circumvent this problem by deforming $\varPhi (\tau )$ to a non-holomorphic 
Poincar\'e series $E (\tau_1 , \tau_2 ; s)$ such that 
\be
\varPhi (\tau ) = \lim_{s\to 0} E(\tau_1 , \tau_2 ; s)\,, 
\ee 
applying the unfolding trick for large $\Re(s)$ and recovering the desired integral 
in the limit $s\to 0$. To illustrate our method, we shall compute the modular integral
$\int \de\mu \varGamma_{(d,d)} \, j(\tau)$ and represent it 
in terms of a `shifted constrained Epstein zeta series', which (unlike the treatment
in \cite{Harvey:1995fq,0919.11036}) makes its invariance
property under ${\rm O} (d,d,\IZ)$ manifest. 

Before closing this introduction, we note that the RSZ method has already been useful
in string theory for studying the distribution of the graded degrees of freedom in tachyon-free oriented closed string vacua, and their connection to the one-loop free 
energy \cite{Angelantonj:2010ic}. Using this method it was shown that non-tachyonic string configurations are characterised by a spectrum of physical excitations that not only must enjoy asymptotic supersymmetry but actually, at very large mass, bosonic and fermionic states are bound to follow a universal oscillating pattern, whose frequencies are related to the non-trivial zeroes of the Riemann $\zeta$-function. Similar studies have then been generalised to higher genus \cite{Cardella:2010bq,Cacciatori:2011fc,Cacciatori:2011qd} where similar constraints on the interactions of physical states are expected to emerge.

Furthermore, we anticipate that further development of the techniques presented in this work will allow to compute a variety of other modular integrals of interest in string theory. In particular, in vacua with broken supersymmetry, it would interesting to study the behaviour of the one-loop effective potential near points of symmetry enhancement, where extra massless states appear. These typically lead to singularities in the vacuum energy that are difficult to analyse using the standard `unfolding method' \cite{kf1,kf2}, whereas they should be fully captured by our new approach. This would allow one to probe the stringy behaviour around the points of enhanced gauge symmetry. It would also be interesting to generalise our methods to higher genus.

The outline of the paper is as follows. In Section \ref{sec_rs}, we recall the original Rankin-Selberg method for functions of rapid decay, and its generalisation to functions of moderate growth by Zagier. Our paraphrasing of \cite{MR656029} is mainly due to the `basic identity' which appears in the body of the proof of \cite{MR656029} which we later require. In Section \ref{sec_lat}, we apply the RSZ method to one-loop modular integrals of symmetric lattice partition functions and recover the celebrated result of \cite{Dixon:1990pc} in very few steps. We further give a proof of a conjecture in \cite{Obers:1999um}, clarifying the relation between the constrained Epstein zeta series of \cite{Obers:1999um} and the Langlands-Eisenstein series studied in \cite{Green:2010wi}. In Section \ref{sec_rapidg}, we develop a variation on the method of Rankin-Selberg and Zagier for functions of rapid growth at the cusp but with finite, or at most power-like divergent, modular integrals. We  apply our procedure to the modular integral of symmetric lattice partition 
functions times the modular $j$-invariant (and its images under the action of the Hecke operators), and we express it in terms of a novel shifted, constrained Epstein Zeta series of ${\rm O} (d,d; \mathbb{Z})$. The appendix collects some properties of Kloosterman sums used in the text.

\section{A brief review of the Rankin-Selberg-Zagier method \label{sec_rs}}

\subsection{Rankin-Selberg method for functions of rapid decay}
Assume that $F(\tau)$ is an automorphic function of the complex variable $\tau=\tau_1+\I\tau_2$ 
 of rapid decay at the cusp, i.e. such that 
$F(\tau)$ vanishes faster that any power of $\tau_2$ at $\tau_2\to\infty$. 
The Rankin-Selberg transform $\Rstar(F,s)$ of $F$ is defined as the Petersson 
product 
\be
\label{defRS}
\Rstar(F,s)\,\equiv\,  \tfrac{1}{2} \zetastar (2s) \, \sum_{\substack{(c,d)\in\IZ,\\ (c,d)=1}} \int_{\cF} \, 
\de\mu\, \frac{\tau_2^s}{ |c\tau +d |^{2s}} \, F(\tau_1 , \tau_2)\,,
\ee
between $F$ and the non-holomorphic Eisenstein series 
\be\label{Edefinition}
 E^\star(\tau ; s) \equiv \zetastar (2s)\,  E(\tau;s)\,,\qquad 
 E(\tau;s) \equiv 
\tfrac{1}{2}\, \sum_{\substack{(c,d)\in \IZ^2,\\ (c,d)=1}}  \frac{\tau_2^s }{|c\, \tau + d|^{2s}} = \sum_{\gamma \in \varGamma_\infty \backslash \varGamma} \left[ \Im \, (\gamma \cdot \tau ) \right]^s
\,.
\ee
Here $\zetastar (s) \equiv \pi^{-s/2}\, \varGamma (s/2) \, \zeta (s)$ is the completed zeta function with simple poles at $s=1,0$ and zeroes in the strip $0< \Re (s) <1$. The integral over the fundamental domain $\cF$ can be unfolded on the strip $\cS=\{\tau_2>0,-\frac12<\tau_1<\frac12\}$,  so that
\be
\label{RSunfold}
\Rstar(F;s) =
\zetastar (2s)\, \int_\cS \frac{\de\tau_1\, \de\tau_2}{\tau_2^{2-s}}\, F(\tau) 
= \zetastar(2s)\, \int_0^\infty \de\tau_2\, \tau_2^{s-2}\, F_0(\tau_2) \,,
\ee
where $F_0(\tau_2)$ is the constant term of $F$,
\be
\label{defF0}
F_0(\tau_2)=\int_{-1/2}^{1/2}  \de\tau_1 \, F(\tau) \, .
\ee 
Thus, $\Rstar(F,s)$ is proportional to the Mellin transform of $F_0$.
Now, $E^\star(\tau;{s})$ is well-known to be a meromorphic function in $s$, with simple poles at $s=0,1$, satisfying the first Kronecker limit formula
\be
\label{kron1}
E^\star(\tau;s) = \frac{1}{2(s-1)} + \tfrac{1}{2} \left( \gamma -  \log ( 4 \pi \, \tau_2 \, |\eta(\tau)|^4) \right)
+ \cO(s-1) \,,
\ee
where $\gamma$ is the Euler-Mascheroni constant and $\eta(\tau)= q^{1/24} \prod_{n=1}^{\infty} (1-q^n)$ is the Dedekind function (as usual, $q=e^{2\pi\I\tau}$). These statements follow from
the Chowla-Selberg formula
  \be
 \label{chowla}
 E^\star(\tau;s) = \zetastar(2s) \, \tau_2^s + \zetastar(2s-1) \, \tau_2^{1-s}
 + 2 \, \sum_{N\neq 0} |N|^{s-\frac12} \sigma_{1-2s}(N) \, \tau_2^{1/2} K_{s-\frac12}(2\pi |N| \tau_2)\,
 e^{2\pi \I N \tau_1}\,,
 \ee
where  $\sigma_t(N)$ is the divisor function
 \be
 \label{defsig}
 \sigma_t(N) \equiv \sum_{0<d|N} d^t \ ,
 \ee
 and $K_t(z)$ is the modified Bessel function of the second kind. The properties  
\be
\label{symxiK}
\zetastar(s)=  \zetastar(1-s) \ ,\qquad K_t(x) = K_{-t}(x)\ ,\qquad  \sigma_t(n) = n^t \, \sigma_{-t}(n)\ ,
\ee
ensure that $E^\star(\tau;s)$ satisfies the functional equation 
 \be
 \label{funcE}
 E^\star(\tau;s)  =  E^\star(\tau;1-s) \ .
 \ee 
It then follows that $\Rstar(F; s)$ inherits the same analytic and functional properties of $ E^\star(\tau;s)$, {\em i.e.} it is a meromorphic function of $s$ with simple poles at $s=0,1$ and symmetric with respect to the critical axis $\Re (s) = \frac{1}{2}$,
\be\label{ferstar}
\Rstar(F; s) = \Rstar(F; 1-s) \,.
\ee
This method was used in the mathematics literature \cite{MR0000411,MR0002626} to establish analytic properties of certain Dirichlet  $L$-series (see e.g. \cite{MR993311} for a survey). More importantly for our purposes, the residue at $s=1$ of the Rankin-Selberg transform is equal to (half) the  average value of $F$ on the fundamental domain $\cF$, 
\be
\label{ResR}
\mathrm{Res}\, \left.  \Rstar(F; s) \right|_{s=1} 
= \tfrac{1}{2} \int_\cF \de\mu\, F  =  -\mathrm{Res} \, \left. \Rstar(F; s) \right|_{s=0}\,.
\ee
This in principle provides a way to evaluate the integral of the automorphic function $F$ on the fundamental domain, from the Mellin transform of the constant  term $F_0$ \cite{Angelantonj:2010ic,Cardella:2010bq,Cacciatori:2011fc}.  In particular, the residue at $s=1$ depends only on the behaviour of $F_0(\tau_2)$ near $\tau_2=0$ .

\subsection{Rankin-Selberg method for functions of moderate growth\label{sec_moderate}}

In physics applications,  $F$ is rarely of rapid decay. Fortunately, the Rankin-Selberg method has been adapted to the case of automorphic functions of moderate growth by Zagier in \cite{MR656029}, which we paraphrase below. 

Let $F(\tau)$ be an automorphic function whose behaviour at the cusp $\tau =\I\infty$ is of the form
\begin{equation}
F(\tau) \sim \varphi (\tau_2) + O (\tau_2^{-N})  \qquad (\forall N>0 ) \,,
\end{equation}
where 
\begin{equation}
\label{defphi}
\varphi (\tau_2) = \sum_{i=1}^\ell \, \frac{c_i }{n_i !} \, \tau_2^{\alpha_i} \, \log^{n_i} \tau_2 
\end{equation}
for suitable $c_i\in\IC, \alpha_i\in\IC, n_i\in\IN$. For this class of functions, following Zagier \cite{MR656029}, we {\it define} the Rankin-Selberg transform as 
\be
\label{defRstar}
\Rstar(F;s) = \zetastar(2s)\, \int_0^\infty \de \tau_2\, \tau_2^{s-2}\,\left(F_0 - \varphi \right)  \ ,
\ee
where $F_0(\tau_2)$ is the $\tau_1$-constant term \eqref{defF0} of $F$. The integral \eqref{defRstar} converges absolutely when $\Re(s)$ is large enough (namely, $\Re(s)>|\Re(\alpha_i)|$ for all $i$). As we shall see\footnote{This statement can be seen right away by noticing that $\varphi(\tau_2)$ is annihilated by  the differential operator $\square\equiv \prod_{i=1}^l[\Delta-\alpha_i(\alpha_i-1)]^{n_i+1}$, where $\Delta$ is the Laplacian on $\cH$, and applying the standard Rankin-Selberg method to the rapidly decaying function $\square F$. We are grateful to D. Zagier for  pointing this out.}, $\Rstar(F;s)$ can be meromorphically continued to all $s$, with possible poles at $s=0,1,\alpha_i$ and $1-\alpha_i$, and is invariant under $s\mapsto 1-s$. Moreover, (half) the residue of  $\Rstar(F;s)$ at $s=1$  gives a  prescription of the  (otherwise divergent)  renormalised integral of $F$ on the fundamental domain. To establish this, we shall use a combination of `hard infrared cut-off' and `zeta function regularisation', {\em i.e.} consider the (manifestly finite) integral
\be
\Rstar_\cT (F; s ) \equiv \int_{\cF_{\cT}}\de\mu\, F(\tau) \, E^\star(\tau;s) 
\ee
on  the ``cut-off fundamental domain" 
$\cF_{\cT} = \cF \cap \{ \tau_2\leq \cT\}$. It is a fundamental domain for the ``cut-off Poincar\'e upper half plane"
\be
\label{cutoffuhp}
\cH_{\cT} = \cH \cap \{ \tau_2\leq \cT\} \ - \bigcup\limits_{\substack{(a,c)\in\IZ^2,\\c\geq 1, (a,c)=1}}
S_{a/c}\ , 
\ee
where $S_{a/c}$ is the disk of radius $1/(2c^2 \cT))$ tangent to the real axis at $a/c$. Defining $\chi_{\cT}$ to be the characteristic function of $\cH_{\cT}$ and performing  the same unfolding trick as in \eqref{RSunfold} with $F\cdot \chi_{\cT}$ in place of $F$, we obtain
\be
\Rstar_\cT (F;s ) 
= \zetastar(2s)\, \int_{\varGamma_\infty\backslash \cH_{\cT}} \de\tau_1 \de\tau_2 \,
F(\tau) \, \tau_2^{s-2}\,  ,
\ee
Using \eqref{cutoffuhp}, this may be rewritten as 
\be 
\label{sumsac}
\Rstar_\cT (F;s) =\zetastar(2s)\, \left( \int_0^{\cT} \de \tau_2\int_{-\frac{1}{2}}^{\frac{1}{2}} \de\tau_1 \,
F(\tau) \, \tau_2^{s-2}\,  
- \sum\limits_{\substack{c\geq 1, (a,c)=1,\\ a \mod c}} \int_{S_{a/c}} \de\tau_1 \de\tau_2\, 
F(\tau) \, \tau_2^{s-2} \right) \,.
\ee
Now, the disc $S_{a/c}$ is mapped to $\cH\cap \{\tau_2>\cT\}$ by any element $\gamma={\scriptstyle \begin{pmatrix}a & b\\c & d\end{pmatrix}}\in\Gamma$. For fixed $a/c$, all such elements are related by right multiplication by $\varGamma_\infty$. Thus, the last term in the bracket in \eqref{sumsac} can be rewritten as 
\be
 \int_{\cF-\cF_{\cT}} \de\mu\, F(\tau) \sum_{\gamma \in\varGamma_\infty\backslash \Gamma \,, c\geq 1}
[\Im(\gamma\cdot \tau)]^{s}\, \de\mu\ .
\ee
The sum over $\gamma$ reproduces the Eisenstein series $E (\tau;s)$, modulo the term $\tau_2^s$ due to the restriction $c\geq 1$. Putting everything together, we find
\be
\label{Zagier0}
\Rstar_\cT (F; s) = 
\zetastar(2s) \, \int_0^{\cT} \de\tau_2\, F_0 (\tau_2 ) \, \tau_2^{s-2}  - \int_{\cF-\cF_{\cT}}\de\mu \, F\, \left( E^\star(\tau;s)- \zetastar(2s)\, \tau_2^s \right) \, .
\ee
The symmetry $s\mapsto 1-s$ may be restored by further subtracting the second constant term from $E^\star(\tau;s)$, so that, after multiplying by $\pi^{-s} \Gamma(s)$ and rearranging terms, 
\be
\begin{split}
& \int_{\cF_{\cT}} \de\mu \, F (\tau_1 ,\tau_2 )\, E^\star(\tau;s) \, \de\mu  +
\int_{\cF-\cF_{\cT}} \de\mu\,F (\tau_1 ,\tau_2 ) \, \left( E^\star(\tau;s) - E^{\star(0)}(\tau;s) \right) 
\\
&\qquad \qquad = \zetastar(2s) \,  \int_0^{\cT}\de \tau_2\,  F_0 (\tau_2 ) \, \tau_2^{s-2}   - 
\zetastar(2s-1)  \int_{\cT}^\infty  \de\tau_2 \, F_0 (\tau_2 ) \, \tau_2^{-1-s} \, ,
\end{split}
\ee
where $E^{\star(0)}(\tau;s) = \zetastar (2s) \, \tau_2^s + \zetastar (2s-1) \, \tau_2^{1-s}$ is the constant term in the Fourier expansion of $E^{\star}(\tau;s)$. Since this is $\tau_1$-independent, the product $F\, E^{\star(0)}(\tau;s)$ appearing in the first line can be replaced by $F_0\, E^{\star(0)}(\tau;s)$ without changing the result of the integral over the strip $\cF-\cF_{\cT}$. Now, the terms in the second line evaluate to
\be
\begin{split}
 \int_0^{\cT} \de\tau_2\,  F_0 (\tau_2 )\, \tau_2^{s-2}   &= \Rstar(F;s)/\zetastar(2s) - 
  \int_{\cT}^\infty \de\tau_2\, (F_0 - \varphi) \, \tau_2^{s-2} + h_{\cT}(s)\,,
\\
 \int_{\cT}^\infty\de\tau_2\,  F_0 (\tau_2 )\, \tau_2^{-1-s} &=
  \int_{\cT}^\infty \de\tau_2\, (F_0 - \varphi) \, \tau_2^{-1-s} + h'_{\cT}(s)\,,
\end{split}
\ee
where $h_{\cT}$ and $h'_{\cT}$ are incomplete Mellin transforms of $\varphi$,
\be
h_{\cT}(s) = \int_0^{\cT} \de\tau_2\, \varphi(\tau_2) \, \tau_2^{s-2}\,,\qquad 
h'_{\cT}(s) = \int_{\cT}^\infty \de\tau_2\, \varphi(\tau_2) \, \tau_2^{-1-s}\, .
\ee
A key fact about the class of functions $\varphi$ in \eqref{defphi} is that their (complete) Mellin transform vanishes, therefore $h'_{\cT}(s)=-h_{\cT}(1-s)$. Moreover,  integrating $\varphi\, \tau_2^{s-2}$ once, 
\be
h_{\cT}(s) = \sum_{i=1}^\ell \frac{c_i}{n_i!}\, \sum_{m=0}^{n_i}\frac{(-1)^{n_i-m}}{m!}\,
\frac{\cT^{s+\alpha_i-1}\, \log^m \, \cT}{(s+\alpha_i-1)^{n_i-m+1}}\,.
\ee
Using this and rearranging terms, we arrive at Zagier's basic identity, Eq. (27) in \cite{MR656029},
\be
\begin{split}
\label{mastereqRS}
\Rstar(F; s) =& \int_{\cF_{\cT}}\de\mu \, F\, E^\star(\tau;s) +
\int_{\cF-\cF_{\cT}} \de\mu \, \left( F\, E^\star(\tau;s) - \varphi\, E^{\star(0)}(\tau;s) \right) 
\\
&-\zetastar(2s) \, h_{\cT}(s) - \zetastar(2s-1)\, h_{\cT}(1-s) \ .
\end{split}
\ee

Evidently, the r.h.s. of \eqref{mastereqRS} is independent of $\cT$, meromorphic  in $s$, invariant under $s\mapsto 1-s$, and analytic away from $s=0,1$  (where $E^\star(\tau;s)$ has a simple pole) and from $s=\alpha_i, 1-\alpha_i$ (where $h_{\cT}(1-s)$, respectively $h_{\cT}(s)$, has a pole of degree $n_i+1$)\footnote{The apparent pole at $s=1/2$ cancels between the last two terms in \eqref{mastereqRS}.}. Thus, when no $\alpha_i$ coincides with $0,1$,
\be
\label{anaRstar}
\Rstar(F;s) = \sum_{i=1}^{l} c_i \left( \frac{\zetastar(2s)}{(1-\alpha_i-s)^{n_i+1}}
+ \frac{\zetastar(2s-1)}{(s-\alpha_i)^{n_i+1}} \right) + \frac{\varPhi(s)}{s(s-1)} \ ,
\ee
where $\varPhi(s)$ is an entire function of $s$. Moreover, the residue at $s=1$ is 
\begin{equation}
\begin{split}
{\rm Res}\left.\,   \Rstar (F ;s) \right|_{s=1} =& - {\rm Res}\,  \left[ \zetastar(2s) \, h_{\cT} (s) \right]_{s=1}
- {\rm Res} \, \left[ \zetastar(2s -1)\, h_{\cT} (1-s) \right]_{s=1} 
\\
&+ \tfrac{1}{2} \left[ \int_{{\mathcal F}_{\cT}} \de\mu\, F(\tau_1,\tau_2 ) + \int_{{\mathcal F}-{\mathcal F}_{\cT}} \de\mu\, \left( F  (\tau_1, \tau_2 ) -\varphi (\tau_2 ) \right)  \right]\,.
\end{split}
\label{Zagier1}
\end{equation}
Letting $\hat \varphi (\tau_2)$ be an anti-derivative of $\varphi (\tau_2)$,
\begin{equation}\label{antiderT}
\hat \varphi (\tau_2) = \sum_{\substack{1\le i \le \ell\\\alpha_i\neq 1}} c_i \, \sum_{m=0}^{n_i} \frac{ (-1)^{n_i -m}}{m!}\, \frac{ \tau_2^{\alpha_i -1}\, \log^m \tau_2}{(\alpha_i -1 )^{n_i -m+1}} 
+ \sum_{\substack{1\le i \le \ell\\ \alpha_i=1}} c_i \, \frac{\log^{n_i +1} \tau_2}{(n_i+1)!}\,,
\end{equation}
and defining the {\em renormalised integral} as 
\begin{equation}
\label{defRN}
\begin{split}
{\rm R.N.} \, \int_{\mathcal F} \de\mu\,  F(\tau_1 , \tau_2)  &= \int_{{\mathcal F}_{\cT}} \de\mu \, F(\tau_1 ,\tau_2 )+ \int_{{\mathcal F}-{\mathcal F}_{\cT}} \de\mu \, \left( F(\tau_1 , \tau_2 ) -\varphi (\tau_2) \right) -\hat \varphi (\cT)
\\
&= \lim_{\cT\to \infty} \left[ \int_{{\mathcal F}_{\cT}} \de\mu \, F(\tau_1 ,\tau_2 )  - \hat \varphi (\cT) \right]\,,
\end{split}
\end{equation}
which is by construction $\cT$-independent, eq. (\ref{Zagier1}) may be rewritten as 
\begin{equation}
{\rm R.N.}\, \int_{\mathcal F} \de\mu\, F(\tau_1 ,\tau_2)  = 2\, {\rm Res} \left[ \Rstar (F;s) + \zetastar (2s)\, h_{\cT} (s) + \zetastar (2s-1) \, h_{\cT} (1-s) \right]_{s=1} -\hat\varphi (\cT)\,.
\label{Zagier2}
\end{equation}
In fact, the r.h.s. of \eqref{mastereqRS} is itself the renormalised integral
\be
\Rstar(F; s) =
{\rm R.N.}\, \int_{\mathcal F} \de\mu\,  F(\tau_1 , \tau_2) \, E^\star(\tau ;s) \, ,
\ee
with $F(\tau_1 , \tau_2) E^\star(s,\tau)$, $\varphi(\tau_2) \, E^{\star(0)}(\tau;s)$ and $\zetastar(2s)\, h_{\cT}(s)+\zetastar(2s-1)\, h_{\cT}(1-s)$ playing the r\^ole of $F(\tau_1 ,\tau_2)$, $\varphi(\tau_2)$ and $\hat\varphi(\cT)$ in \eqref{defRN}, respectively.

For functions of rapid decay, the renormalised integral reduces to the usual integral and $h_{\cT}, \hat\varphi(\cT)$ vanish, hence \eqref{Zagier2} reduces to \eqref{ResR}. More generally, if $\Re(\alpha_i)<1$ for all $i$, the integral $\int_{\cF} \de \mu\, F$ still converges, and one can take the limit $\cT\to\infty$ in  \eqref{Zagier2} and recover \eqref{ResR}. If however one of the $\Re(\alpha_i)\geq 1$,  the integral  $\int_{\cF_{\cT}} F \,\de\mu$ diverges like $\hat\varphi(\cT)$ as a function of the infrared cut-off, and  \eqref{Zagier2} provides a renormalisation prescription which depends only on the divergent terms with $\Re(\alpha_i)\geq 1$ in \eqref{defphi}.

Of course, the renormalisation prescription \eqref{defRN} is by no means the only possible one. For example, one may decide to subtract the non-decaying  part from $F$ and integrate the remainder on the fundamental domain, defining
\begin{equation}
\label{defRNDKL}
\begin{split}
{\rm R.N.}' \, \int_{\mathcal F} \de\mu \, F(\tau_1 , \tau_2 ) &= \int_{{\mathcal F}} \de\mu\,  \left( F(\tau_1 , \tau_2) 
-\varphi([\tau_2]) \right)\,.
\end{split}
\end{equation}
Here we denoted by $[\tau_2]$ the imaginary part of $\gamma\cdot\tau$, where $\gamma$ is an element of $\Gamma$ which maps $\tau$ into the standard fundamental domain $\cF$ (so $[\tau_2]=\tau_2$ if $\tau\in \cF$). The renormalised integrals \eqref{defRNDKL} and \eqref{defRN} differ by a finite quantity
\be
\varDelta \equiv {\rm R.N.}\, \int_{\mathcal F} \de\mu\, F(\tau_1,\tau_2) - {\rm R.N.}' \, \int_{\mathcal F} \de\mu\, F(\tau_1 , \tau_2) =
 \lim_{\cT\to \infty} \left[ \int_{{\mathcal F}_{\cT}} \de\mu\, \varphi([\tau_2]) - \hat \varphi (\cT) \right]\,,
\ee 
which can be computed explicitly. For example, if all the $n_i$'s are zero, 
\be
\label{eqdelta}
\varDelta = \sum_{i=1}^\ell c_i\, \frac{\delta_{\alpha_i ,1}-{}_2 F_1 (\frac{1}{2},\frac{1-\alpha_i}{2};\frac{3}{2};\frac{1}{4})}{\alpha_i -1} \,,
\ee
with $\delta_{a,b}$ being the Kronecker symbol, and  $_2F_1 (a,b;c;z)$ the standard hypergeometric function. Notice that the case $\alpha_i =1$ is well encoded in \eqref{eqdelta} since, in the limit $\alpha_i \to 1$, 
\be
{}_2 F_1 \left(\frac{1}{2},\frac{1-\alpha_i}{2};\frac{3}{2};\frac{1}{4}\right) = 1 - \left(1-\log \frac{3\sqrt{3}}{2} \right) \, (\alpha_i -1) + \cO ((\alpha_i -1)^2)\,,
\ee
and thus $\varDelta = 1-\log \,\frac{3\sqrt{3}}{2}$.
 
\section{Lattice modular integrals and constrained Epstein zeta series \label{sec_lat}}

In this section, we apply the Rankin-Selberg method to the evaluation of the integral of the lattice partition function\footnote{Here and in the following we set $\alpha ' =1$, and we suppress the explicit dependence of the lattice on $\tau_1$ and $\tau_2$.} 
\begin{equation}
\label{gdd}
\varGamma_{(d,d)} (g,B) = \tau_2^{d/2}\, \sum q^{\frac{1}{2}\, p_{L\, i } g^{ij} p_{ L\, j } }\, \bar q ^{\frac{1}{2}\, p_{R\, i} g^{ij} p_{R\, j }} =
\tau_2^{d/2} \, \sum_{(m_i,n^i)\in \IZ^{2d}} e^{-\pi \tau_2\, {\mathcal M}^2} \, e^{2\pi\I \tau_1 \, m _i \,  n^i}
\end{equation}
on the fundamental domain. The left-handed and right-handed momenta
\be
p_{L(R)\, i} = \frac{1}{\sqrt{2}}\left(\, m_j \pm (g_{ij} \mp B_{ij} ) \, n^j \,\right) \,,
\ee
depend on the geometric data of the compactification torus, {\em i.e.} the metric $g_{ij}$ of the $T^d$
and the NS-NS two-form $B_{ij}$, that together parameterise the symmetric space ${\rm O}(d,d)/{\rm O} (d)\times {\rm O}(d)$, also known as the Narain moduli space. 
In the last equality of \eqref{gdd},
\be
{\mathcal M}^2= (m_i+B_{ik} n^k) g^{ij} (m_j+B_{jl} n^l) + n^i g_{ij} n^j=p_L^2+p_R^2
\ee
is the mass-squared of a string ground state with Kaluza-Klein momentum $m_i$ and winding number $n^i$, with $g^{ij}$  being the inverse metric. The  lattice partition function is manifestly invariant under  ${\rm O} (d,d;\IZ)$, but also under ${\rm SL}(2;\IZ)_{\tau}$. This invariance  is exposed after Poisson resummation with respect to $m_i$,
\be
\varGamma_{(d,d)} (g,B ) = \sqrt{{\rm det}\, g}\,
\sum_{(m^i,n^i)\in\IZ^{2d}} \exp\left[ -\pi\ \frac{(m^i+n^i\tau)\, g_{ij}
(m^j+n^j\bar\tau)}{\tau_2}+2\pi \I\, B_{ij} \,m^i n^j\right] \,.
\ee
The partition function for the Narain lattice clearly belongs to the class of functions considered in Section \ref{sec_moderate} since, in the limit $\tau_2 \to \infty$,
\be
\varGamma_{(d,d)} (g,B ) \sim  \tau_2^{d/2} = \varphi (\tau_2 ) \,,
\ee 
that matches eq. \eqref{defphi} with with $\ell=1$, $\alpha_i=d/2$ and $n_i=0$. Using Zagier's extension of the Rankin-Selberg method, as summarised in the previous section, we can then compute the renormalised integral
\be
\label{RGamma}
\Rstar(\varGamma_{(d,d)}; s) =
{\rm R.N.}\, \int_{\mathcal F} \de\mu \, \varGamma_{(d,d)} (g,B) \, E^\star(s,\tau) \,, 
\ee
and especially its residue at $s=0$, which is proportional to the IR finite one-loop integral of
the lattice partition function,
\be
I_d (g,B) = {\rm R.N.}\, \int_{\mathcal F}\de\mu\,  \varGamma_{(d,d)} (g,B ) \,.
\ee

Before proceeding with the explicit calculation of the integral (\ref{RGamma}), we notice that the Rankin-Selberg transform $\Rstar(\varGamma_{(d,d)}, s) $ is an eigenfunction of the Laplace operator acting on the moduli space ${\rm O} (d,d)/{\rm O} (d)\times {\rm O} (d)$
\be\label{Rstarde}
\Delta_{{\rm SO} (d,d)}\, \Rstar(\varGamma_{(d,d)}, s) = \tfrac{1}{4}\,  (2s-d)(2s+d-2)\, \Rstar(\varGamma_{(d,d)}, s)\,,
\ee
with 
\be\label{DelSOR}
\Delta_{{\rm SO} (d,d)} = \tfrac{1}{4}\, g^{ik}\, g^{jl} \left( \frac{\partial}{\partial g_{ij}}\, 
 \frac{\partial}{\partial g_{kl}} +  \frac{\partial}{\partial B_{ij}}\, 
 \frac{\partial}{\partial B_{kl}} \right) + \tfrac{1}{2} \, g^{ij}\, \frac{\partial}{\partial g_{ij}}
 \,.
 \ee
This follows straightforwardly from the following differential equations satisfied by the partition function of the Narain lattice \cite{Obers:1999um} and by the Eisenstein series
\begin{subequations}
\be
\label{DiffEqEisensteina}
0= \left[\Delta_{{\rm SO}(d,d)} - 2 \, \Delta_{{\rm SL} (2)}+\tfrac{1}{4}\, d(d-2) \right]\, \varGamma_{(d,d)} (g,B) \,,
\ee
\be
\label{DiffEqEisensteinb}
0 = \left[ \Delta_{{\rm SL} (2)} -\tfrac{1}{2}\, s(s-1) \right]\, E^\star (\tau ; s)  \,,
\ee
\end{subequations}
where
\be
\label{DeltaSL2}
\Delta_{{\rm SL}(2)}= \tfrac{1}{2}\, \tau_2^2 \, \left( \frac{\partial^2}{\partial \tau_1^2} +  \frac{\partial^2}{\partial \tau_2^2} \right)
\ee
is the Laplace operator on the hyperbolic plane.

\subsection{Constrained Epstein zeta series in dimension $d>2$}

For a generic $d$-dimensional lattice \eqref{gdd},
 \begin{equation}
h_{\cT} (s) = \frac{\cT^{s+d/2-1}}{s+\tfrac{1}{2}\, d -1} \,,
\qquad 
\hat\varphi (\tau_2 ) = \left\{ 
\begin{array}{cl} 
\tau_2^{d/2-1}/( \tfrac{1}{2}\, d-1) &
\mbox{if}\ d\neq 2 \,,
\\
\log\, \tau_2 & \mbox{if}\ d=2\,,
\end{array}
\right.
\end{equation}
and as expected, the integral 
\be
\int_{\mathcal F} \de\mu\, \varGamma_{(d,d)}(g,B)\, E^\star(s,\tau) 
\ee
is power-divergent, logarithmically divergent and absolutely convergent for $s+\tfrac{1}{2}\, d-1>0$, 
$s+\tfrac{1}{2}\, d-1=0$ and  $s+ \tfrac{1}{2}\, d -1<0$, respectively. In all cases, however, the renormalised integral \eqref{RGamma} is finite and is given by the Mellin transform 
\be
\label{REps}
\begin{split}
\Rstar (\varGamma_{(d,d)};s ) &= \zetastar (2s) \, \int_0^\infty  \de\tau_2\, 
\tau_2^{s+d/2-2} \, {\sum_{m_i,n^i}}^\prime\,  e^{-\pi \tau_2\, {\mathcal M}^2} \, e^{2\pi\I\tau_1 \, m\cdot n}
\\
&= \frac{\zetastar (2s) \, \varGamma (s+\tfrac{1}{2}\, d-1)}{\pi^{s+d/2-1}} \, \mathcal{E}^d_V (g,B;s+\tfrac{1}{2}\, d-1)\,.
\end{split}
\ee
Here, $m\cdot n = m_i \, n^i$, and we have denoted by $\mathcal{E}^d_V (g,B;s)$ the constrained Epstein zeta series in the vectorial
representation of ${\rm O} (d,d;\IZ)$, introduced in \cite{Obers:1999um}
\begin{equation}
\label{defceps}
\mathcal{E}^d_V (g,B;s) \equiv {\sum_{m,n}}^\prime \,\frac{\delta (m\cdot n)}{{\mathcal M}^{2s}}
\,,
\end{equation}
which converges absolutely for $s>d$ (as usual, a primed sum does not involve the contribution from $m_i=n^i=0$). 
It is useful to define the {\em completed constrained Epstein zeta series}
\be
\label{compEps}
{\mathcal E}^{d\, \star}_V (g,B;s) \equiv \pi^{-s}\, \varGamma(s)\, \zetastar(2s-d+2) \,  
{\mathcal E}^d_V (g,B;s)\,,
\ee
so that 
 \be
 \Rstar (\varGamma_{(d,d)};s ) = {\mathcal E}^{d\, \star}_V (g,B;s+\tfrac{1}{2}\, d-1)\,.
 \ee
 From eq. \eqref{ferstar} it follows that ${\mathcal E}^{d\, \star}_V (g,B;s)$ satisfies the functional equation
 \be
 \label{funodd}
 {\mathcal E}^{d\star}_V (g,B;s) =   {\mathcal E}^{d\,\star}_V (g,B;d-1-s)\,,
 \ee
 in agreement with eq. \eqref{Rstarde}. 
 
 These properties, together with the invariance of ${\mathcal E}^{d\star}_V (g,B;s)$   under the ring of ${\rm O} (d,d)$-invariant differential operators  \cite{Obers:1999um}, implies that the constrained Epstein zeta series coincides with  the degenerate Langlands-Eisenstein series for ${\rm O}(d,d)$ based on the parabolic subgroup $P$ with Levi subgroup $\IR^+\times {\rm SO}(d-1,d-1)$ \cite{Green:2010wi,Pioline:2010kb,Green:2010kv}. Moreover, from the general statement below \eqref{mastereqRS}, it follows that,   for $d>2$, ${\mathcal E}^{d\star}_V (g,B;s)$ has simple poles at $s=0,d/2-1,d/2$ and $1$, as indicated in \cite{Pioline:2010kb}. 
  
Using the functional equation \eqref{funodd}, we see that \eqref{REps} is equivalent to
 \be
{\rm R.N.}\, \int_{\mathcal F} \de\mu\,  \varGamma_{(d,d)} (g,B ) \, E^\star(s,\tau) = 
{\mathcal E}^{d\, \star}_V \left(g,B;\tfrac{1}{2}\, d-s\right) \,.
 \ee
For $d>2$ one can easily extract the residue at $s=1$ to get 
 \be\label{Idb2}
 \begin{split}
 I_d &= \frac{\Gamma(d/2-1)}{\pi^{d/2-1}}\ {\mathcal E}^d_V \left(g,B;\tfrac{1}{2}\, d-1\right)
 \\
 &= \frac{\pi}{3}\, \frac{ \Gamma(d/2)}{\pi^{d/2}} \, {\rm Res}\left.  {\mathcal E}^d_V \left( g,B;s+\tfrac{1}{2}\, d-1\right) \right|_{s=1} \,,
 \end{split}
 \ee
where in writing the last expression we have made use of the functional equation \eqref{funodd}. 

Notice that the first equality in \eqref{Idb2} establishes Theorem 4 in \cite{Obers:1999um} rigorously. Moreover, comparing with Eq. (C.2) in \cite{Green:2010wi} (and dropping the superfluous volume subtraction), we recognise that the constrained Epstein zeta series is in fact  equal to 
 \be
 {\mathcal E}^{d}_V (g,B;s) =E_{[1,0^{d-1}];s}^{{\rm SO}(d,d)}\,,
 \ee
 where $E_{[1,0^{d-1}];s}^{{\rm SO}(d,d)}$ is the Langlands-Eisenstein series, introduced in \cite{Green:2010wi}. This relation can also be checked by comparing the large volume expansions given in Eq. (C.7) in \cite{Green:2010wi} and in Appendix C.1 of \cite{Obers:1999um}.

\subsection{Low dimension \label{sec_lowdim}}

The cases $d\le 2$ are special, since the integrals are at most logarithmically divergent, and the delta-function constraints in the definitions of the constrained Epstein zeta functions can be explicitly solved.

For $d=1$,  the  lattice partition function reduces to 
\begin{equation}
\label{g11}
\varGamma_{(1,1)} (R) = \sqrt{\tau_2}\,  \sum_{m,n} e^{-\pi \tau_2 \left[ (m/R)^2 + (nR)^2 \right]}\, e^{2i\pi\tau_1 \, mn}\,,
\end{equation}
with $\varphi (\tau_2 ) = \sqrt{\tau_2}$. The modular integral $I_1$ is finite and coincides with the renormalised one. The constrained Epstein zeta series \eqref{defceps} evaluates to 
\be
 {\mathcal E}^{1}_V (g,B;s) = 2\, \zeta(2s) \, ( R^{2s} + R^{-2s} )\,,
\ee
and thus
\be
\Rstar (\varGamma_{(1,1)};s) =  {\mathcal E}^{1,\star}_V (g,B;s-\tfrac{1}{2}) 
= 2 \, \zetastar(2s)\, \zetastar(2s-1) \, \left( R^{1-2s} + R^{2s-1} \right) \,.
\ee
The r.h.s. has simple poles at $s=0,1$ and a double pole at $s=1/2$, in agreement with the general statement below \eqref{mastereqRS}. The residue at $s=1$ produces then the standard result for the modular integral
\be\label{1dlattice}
I_1 = \int_{\mathcal F} \de\mu\, \varGamma_{(1,1)} (R) = \frac{\pi}{3}\, \left( R+R^{-1} \right)\,,
\end{equation}
that is invariant under  $R\mapsto 1/R$ as required by the ${\rm O}(1,1;\IZ)$ symmetry of the lattice. Clearly, the same result arises by unfolding the sum over $m,n$ in \eqref{g11}.

For $d=2$, the modular integral $\int_{\cF_{\cT}} \de\mu\, \varGamma_{(2,2)}$ is logarithmically divergent as $\cT\to\infty$. The Rankin-Selberg transform is however finite for large $\Re(s)$ and still given by the constrained Epstein zeta series \eqref{defceps}. Once more, the constraint can be explicitly solved to get the standard expression for the integral. To this end, it is convenient to parameterise the two-torus in terms of the complex structure modulus $U=U_1 + i U_2$ and K\"ahler modulus  
$T=T_1 + i T_2$, so that 
\be
p_L= \frac{m_1 +U m_2+\bar T(n^2-U n^1)}{\sqrt{ 2T_2 U_2}}
\ ,\quad p_R=\frac{m_1 +U m_2+T(n^2-U n^1)}{\sqrt{ 2T_2 U_2}}\ .
\ee
Now, we notice that the most general solution of the constraint $m_1 n^1+m_2 n^2=0$ is given by the elements in
the disjoint union
\be
	(m_1,m_2,  n^1,n^2) \in \mathcal{S}_1\cup\mathcal{S}_2~,
\ee
with $\mathcal{S}_1, \mathcal{S}_2$ being the sets:
\be\label{DiophantineSol}
\begin{split}
&\mathcal{S}_1 =   \{  (m_1,m_2,0,0) \ ,(m_1,m_2)\in \IZ^2 \}~, \\
&\mathcal{S}_2 = \{ (c \tilde m_1, c \tilde m_2, -d \tilde m_2, d \tilde m_1)\ ,
(\tilde m_1,\tilde m_2)\in \IZ^2\ , \gcd(\tilde m_1,\tilde m_2)=1\ , (c,d)\in \IZ\ ,d\geq 1\}~.
\end{split}
\ee
The contribution of the first set to ${\mathcal E}^{2\star}_V (T,U;s)$ easily gives
\be
\zetastar(2s)\sum\limits_{\substack{ (m_1,m_2)\in\mathbb{Z}^2\\ (m_1,n_1)\neq(0,0) }}{~\left[\frac{T_2 U_2}{|m_1+Um_2|^2}\right]^s} = 2T_2^s\, \zeta(2s)\,E^\star (U;s)~,
 \ee
 where $E^\star(\tau;s)$ is the ${\rm SL}(2,\IZ)$ invariant Eisenstein series (\ref{Edefinition}). 
For solutions belonging to the second set, the mass-squared factorises as 
\be
\cM^2 = |p_L|^2+|p_R|^2 = \frac{|\tilde m_1 + U \tilde m_2|^2}{U_2}\times
\frac{|c + T d|^2}{T_2}\,, 
\ee
so that the contribution of  $\mathcal{S}_2$ gives
\be
 \zetastar(2s)\sum\limits_{\substack{ (c,d)\in\mathbb{Z}^2\\ d\geq 1 }}{~\left[\frac{T_2}{|c+T d|^2}\right]^s}  \sum\limits_{\substack{ (\tilde{m}_1,\tilde{m}_2)\in\mathbb{Z}^2\\ (\tilde{m}_1,\tilde{m}_2)= 1 }}{~\left[\frac{U_2}{|\tilde{m}_1+U \tilde{m}_2|^2}\right]^s} = 2 E^{\star}(U;s) \sum\limits_{\substack{ (c,d)\in\mathbb{Z}^2\\ d\geq 1 }}{~\left[\frac{T_2}{|c+T d|^2}\right]^s}~.
\ee
These two results combine into a simple expression for  the constrained  Epstein zeta series, 
\be
\label{E2}
 \Rstar (\varGamma_{(2,2)} ;s )  =
 {\mathcal E}^{2\star}_V (T,U;s) = 2\,  E^\star (T;s) \, E^\star (U;s)\, .
\ee
This relation is actually a consequence of the group isomorphism 
\be\label{O22iso}
{\rm O} (2,2; \IZ) \sim {\rm SL} (2;\IZ)_T\times {\rm SL}(2;\IZ)_U \ltimes \IZ_2\,, 
\ee
and of the decomposition of the vectorial representation of  ${\rm O} (2,2)$ in terms of the bi-fundamental $(2,2)$ of ${\rm SL} (2) \times {\rm SL} (2)$. 

It is clear from \eqref{E2} that ${\mathcal E}^{2\star}_V (T,U,s)$ has a double pole at $s=0$ and 
$s=1$. Once more, this is in agreement with \eqref{mastereqRS}, since, upon using 
$\zetastar(s)=1/(s-1)+\frac12(\gamma-\log4\pi)+\cO(s-1)$ and $h(\cT)=\cT^s/s$, 
the  second line in this equation has a double
pole at $s=0,1$,
\begin{equation}
\label{zeht2}
\zetastar (2s) \, h_{\cT} (s) + \zetastar (2s-1) \, h_{\cT} (1-s)  = -\frac{1}{2(s-1)^2}+
 \frac{1}{2(s-1)} \left[\log (4 \pi \, \cT) -\gamma \right] + \dots~~.
\end{equation}
To compute the renormalised one-loop integral $I_2$, one then needs to extract the residue at $s=1$. Using  the first Kronecker limit formula \eqref{kron1},  one finds 
\be
\label{zeht3}
 \Rstar (\varGamma_{(2,2)} ;s )  =
\frac{1}{2(s-1)^2} + \frac{1}{s-1}
\left[
  \gamma - \tfrac{1}{2}\, \log \left( 16 \pi^2 \, T_2 \, U_2 \, |\eta (T) \, \eta (U) |^4 \right) \right]
  +\dots  \,.
\ee
Combining  \eqref{zeht2},  \eqref{zeht3} with  \eqref{Zagier2} 
and using $\hat\varphi(\cT)=\log(\cT)$, one arrives at the following expression for the {\em renormalised integral} of the two-dimensional Narain lattice partition
function
\begin{equation}
\label{I2}
I_2 = {\rm R.N.}\, \int_{{\mathcal F}} \varGamma_{(2,2)} (T,U ) \, d\mu = - \log \left( 4\pi \, e^{-\gamma}\, T_2 \, U_2 \, |\eta (T)\, \eta (U) |^4 \right)\,.
\end{equation}
We stress that Eq. \eqref{I2} is the result of the 
renormalisation prescription \eqref{defRN}. It differs by a 
finite constant from the renormalised integral computed in \cite{Dixon:1990pc}
using the renormalisation prescription \eqref{defRNDKL}. Indeed, using \eqref{eqdelta}
one arrives at 
\be
\label{dkl}
\int_{{\mathcal F}} \left(\varGamma_{(2,2)} (T,U)-\tau_2 \right) \, d\mu 
= - \log \left( \frac{8\pi e^{1-\gamma}}{3\sqrt{3}} \, T_2 \, U_2 \, |\eta (T)\, \eta (U) |^4 \right)\, ,
\ee
in agreement with \cite{Dixon:1990pc}.  In our opinion, the derivation of 
\eqref{dkl} via the Rankin-Selberg-Zagier method is considerably simpler than the original
derivation in \cite{Dixon:1990pc}. 

Finally, we comment on the properties of the Rankin-Selberg transform \eqref{E2} and 
the renormalised integral \eqref{I2} of the lattice partition function $\varGamma_{(2,2)}$
under the action of T-duality invariant differential operators. For $d=2$, Eq. 
\eqref{DiffEqEisensteina} becomes 
\be
\label{Diff2}
\left[ -2 \Delta_{{\rm SL}(2)}^{(\tau)} + \Delta_{{\rm SL}(2)}^{(T)} 
+  \Delta_{{\rm SL}(2)}^{(U)} \right]\, \varGamma_{(2,2)} =  0\ ,
\ee
where $\Delta_{{\rm SL}(2)}^{(T,U)}$ are the analogue of the Laplacian \eqref{DeltaSL2} 
acting on the hyperbolic $T,U$-plane. However, one may check that $\varGamma_{(2,2)}$
satisfies the stronger equations
\be
\label{Diff2s}
\Delta_{{\rm SL}(2)}^{(\tau)} \, \varGamma_{(2,2)}  = 
\Delta_{{\rm SL}(2)}^{(T)} \, \, \varGamma_{(2,2)} =
\Delta_{{\rm SL}(2)}^{(U)} \, \varGamma_{(2,2)} \ .
\ee
Combining these equations with \eqref{DiffEqEisensteinb}, we
find that $\Rstar (\varGamma_{(2,2)} ;s )$ must be an eigenmode of 
$\Delta_{{\rm SL}(2)}^{(T)}$ and $\Delta_{{\rm SL}(2)}^{(U)}$ separately, 
with the same eigenvalue $\tfrac12 s(s-1)$. This is of course manifest from
the explicit result \eqref{E2}. Moreover, due to the subtraction of the second
order pole in \eqref{zeht3}, the renormalised integral is a quasi-harmonic function 
of the moduli, namely it satisfies 
\be
 \Delta_{{\rm SL}(2)}^{(T)}\, I_2  = \Delta_{{\rm SL}(2)}^{(U)} \, I_2 = \tfrac{1}{2}\ .
\ee
This is again manifest from the explicit result \eqref{I2}. These considerations
will become fruitful when computing integrals with unphysical tachyons in Section \ref{homlin}.

\subsection{Decompactification}

It is useful to ask about the behaviour of the modular integral \eqref{RGamma}
when the radius $R$ of one circle in the $d$-dimensional torus is sent to 
infinity. For simplicity, we restrict to the subspace of the Narain moduli space where 
the lattice partition function factorises into
\be
\varGamma_{(d,d)} (g,B) =
 \varGamma_{(1,1)} (R) \times \varGamma_{(d-1,d-1)} (\tilde g,\tilde B)\,,
\ee
where  $\tilde g$, $\tilde B$ are the metric and NS-NS two-form on the remaining $(d-1)$-dimensional torus. To investigate the limit $R\to\infty$, we consider the modular 
integral with a hard cut-off, and unfold the lattice sum  $\varGamma_{(1,1)}$.
Following the same reasoning as in eqs. \eqref{sumsac}-\eqref{Zagier0} 
one finds, for large enough $\Re(s)$,
\be
\begin{split}
\label{dec1}
\cR_\cT^\star(\varGamma_{(d,d)};s)=& 
R\, \int_{\cF_\cT} \de\mu \,  \varGamma_{(d-1,d-1)} (\tilde g,\tilde B;\tau ) \, E^\star(\tau;s) \\
&+\int_0^\cT  \int_{-1/2}^{1/2} \frac{\de\tau_1\de\tau_2}{\tau_2^{2}}\,
\left( \sum_{m\neq 0} R\, e^{- \frac{\pi R^2 m^2}{\tau_2}} \right)\, 
\varGamma_{(d-1,d-1)} (\tilde g,\tilde B;\tau )\, E^\star(\tau;s)  \\
&- \int_\cT^\infty  \int_{-1/2}^{1/2} \frac{\de\tau_1\de\tau_2}{\tau_2^{2}}\,
\left( \sum_{m\in\IZ,n\neq 0}R\, e^{-\frac{\pi R^2 |m-n\tau|^2}{\tau_2}} \right)\, 
\varGamma_{(d-1,d-1)} (\tilde g,\tilde B;\tau )\, E^\star(\tau;s)  \,.
\end{split}
\ee
In the limit $R\to\infty$, the second and third lines are exponentially suppressed, except for the contribution of the massless sector of $\varGamma_{(d-1,d-1)}$ and
the zero-mode part of $E^\star(\tau;s)$ to the second line, which may  be replaced by
\be
\label{dec2}
\begin{split}
\int_0^\cT  \int_{-1/2}^{1/2} \, \de\mu \,
\left( \sum_{m\neq 0} R\, e^{- \frac{\pi R^2 m^2}{\tau_2}} \right)\, 
\left[ \zetastar(2s)\, \tau_2^{s+\frac{d-1}{2}} +  \zetastar(2s-1)\, \tau_2^{1-s+\frac{d-1}{2}} 
\right]\ .
\end{split}
\ee
The renormalised integral $\cR^\star(\varGamma_{(d,d)};s)$ may therefore be written as 
\be
\begin{split}
\label{dec3}
\cR^\star(\varGamma_{(d,d)};s)=&  R\, \cR^\star(\varGamma_{(d-1,d-1)};s) \\
&+ 2 \, \zetastar(2s)\, \lim_{\cT\to\infty}
\left[ 
\int_0^\cT  \int_{-1/2}^{1/2} \, \de\mu\, 
 \tau_2^{s+\frac{d-1}{2}}  \left( \sum_{m\in\IZ} R\, e^{- \frac{\pi R^2 m^2}{\tau_2}} 
 -\tau_2^{1/2} \right) \right]\\
 &+ 2  \, \zetastar(2s-1)\, \lim_{\cT\to\infty}
\left[ 
\int_0^\cT  \int_{-1/2}^{1/2} \, \de\mu\, \tau_2^{\frac{d+1}{2}-s}  
 \left( \sum_{m\in\IZ} R\, e^{- \frac{\pi R^2 m^2}{\tau_2}} 
 -\tau_2^{1/2} \right) \right]\\
 &+\dots\,,
\end{split}
\ee
where the ellipses denote exponentially suppressed corrections in the limit $R\to\infty$.
A Poisson resummation over $m$ shows that the term in round brackets is exponentially suppressed as $\tau_2\to\infty$, and the limit $\cT\to\infty$ is therefore finite, and given by 
\be
\label{dec4}
\begin{split}
\cR^\star(\varGamma_{(d,d)};s)=& 
R\, \cR^\star(\varGamma_{(d-1,d-1)}; s) \\ + & 2 
\, \zetastar(2s)\, \zetastar(2s+d-2) \, R^{2s+d-2} + 2\,\zetastar(2s-1)\,  \zetastar(2s-d+1)\, 
R^{d-2s}+\dots\ ,
\end{split}
\ee
up to exponentially suppressed corrections as $R\to\infty$.
This formula is manifestly invariant under $s\mapsto 1-s$, and provides the constant term for
the constrained Epstein zeta series \eqref{compEps} with respect to the parabolic 
subgroup with Levi component ${\rm SO}(d-1,d-1)\times \IR^+$ inside ${\rm SO}(d,d)$. It is easy
to check that it is satisfied for $d=1,2$ using the explicit results in the previous subsection,
and that it agrees with Eq. (D.16) in \cite{Green:2010wi} for $d=3$. One may also check that 
\eqref{dec3} is consistent with the Laplace equation \eqref{Rstarde} 
using Eq. (A.30) in \cite{Obers:1999um}.

\subsection{Another modular invariant regulator}

In the context of threshold corrections in four-dimensional heterotic vacua, 
Kiritsis and Kounnas \cite{Kiritsis:1994ta} have proposed a different 
modular invariant regularisation for 
on shell infrared divergences, based on replacing Minkowski
space by a superconformal field theory with the same central charge, but 
depending on an infrared cut-off $\varLambda$. For non-zero value of $\varLambda$, 
the threshold correction is given by a modular integral
\be
\label{IKK}
I_d^{KK} (g,B) =\int_{\mathcal F}\de\mu\,  
\varGamma_{(d,d)} (g,B;\tau ) \, Z(\varLambda,\tau)\,,
\ee
where $Z(\varLambda,\tau)$ is related to the partition function $\varGamma_{(1,1)} (R,\tau )$
of a compact boson of radius $R=1/\varLambda$ by
\be
Z(\varLambda,\tau) = D\cdot \varGamma_{(1,1)} (1/\varLambda,\tau )\ ,\qquad 
[D\cdot f](\varLambda) \equiv 2\, \Lambda^2 \partial_\varLambda\, \left[ f(2\varLambda)-f(\varLambda) \right]\,.
\ee
The integral is manifestly finite, since the cut-off function $Z(\varLambda,\tau)$ decays
exponentially at $\tau_2\to\infty$.  For fixed $\tau$, the integrand in
\eqref{IKK} agrees with the usual one in the limit $\varLambda\to 0$, since 
$Z(\varLambda\to 0,\tau)=1$ up to exponential corrections. 

To relate this prescription to ours, we apply the Ranking-Selberg method for functions of rapid decay, and write
\be
I_d^{KK} (g,B) = 2\, {\rm Res}_{s=1} \, \Rstar(\varGamma_{(d,d)} Z(\varLambda,\tau)\ , s) \,,
\ee
where 
\be
\label{RGammaKK}
\Rstar(\varGamma_{(d,d)} Z(\varLambda,\tau)\ , s) = {\rm R.N.}\, 
\int_{\mathcal F} \de\mu \, \varGamma_{(d,d)} (g,B;\tau ) \, Z(\varLambda,\tau)\, E^\star(s,\tau) \ .
\ee
Since $Z(\varLambda,\tau)$ is of rapid decay, the sign ${\rm R.N.}$ is superfluous, 
however introducing it allows us to take the operator $D$ out of the integral and obtain
\be
\label{RGammaKK2}
\Rstar(\varGamma_{(d,d)} Z(\varLambda,\tau)\ , s) = D\cdot \left[ {\rm R.N.}\, 
\int_{\mathcal F} \de\mu \, \varGamma_{(d,d)} (g,B;\tau ) \, 
\varGamma_{(1,1)}(1/\varLambda)\, E^\star(s,\tau) \right] \ .
\ee
Using the results of the previous subsection (with $d,R$ replaced by $d+1, 1/\varLambda$),
we find that in the limit where the infrared cut-off $\varLambda$ is removed, 
\be
\begin{split}
\Rstar(\varGamma_{(d,d)} Z(\varLambda,\tau)\ , s) & \stackrel{\varLambda\to 0} {\longrightarrow}
 \cR^\star(\varGamma_{(d,d)}; s) \\ & + 2 
\, \zetastar(2s)\, \zetastar(2s+d-1) \, D\cdot \varLambda^{1-d-2s} 
\\
&+ 2\,\zetastar(2s-1)\,  \zetastar(2s-d)\, 
D\cdot \varLambda^{2s-d-1} \,.
\end{split}
\ee
Extracting the residue at $s=1$, we find that for $d=0,1$, \eqref{IKK} is finite in the limit 
$\varLambda\to 0$ and reduces to the standard results. 
For $d>2$, \eqref{IKK} 
agrees with \eqref{Idb2} after subtracting the order $\cO(\varLambda^{2-d})$ divergent term.
Finally, for $d=2$ \eqref{IKK} diverges logarithmically as $\varLambda\to 0$,
and agrees with \eqref{I2} after subtracting $-2\log(2 {\rm e} \varLambda)$.  

\section{Modular integrals with unphysical tachyons \label{sec_rapidg}}

In heterotic string vacua with a `spectator' $d$-dimensional torus $T^d$, an interesting class of couplings  in the low energy effective action is determined by the one-loop integral of the elliptic genus \cite{Lerche:1987qk}, 
\be
\label{inthet}
\int_{\mathcal F} \varGamma_{(d+k,d)} (g,B,y )\, \varPhi(\tau)\,\de\mu \,,
\ee
where
\be
\varGamma_{(d+k,d)}(g,B,y )\ = \tau_2^{d/2}\, 
\sum_{p_L,p_R} q^{\frac12 p_L^2}\, \bar q^{\frac12 p_R^2}\
\ee
 is the partition function of the Narain lattice, parameterised by the metric $g_{ij}$, two-form $B_{ij}$ and by Wilson lines $y_i^a$ ($a=1\dots k$), and $\varPhi (\tau)$ is a weak holomorphic 
 modular form of negative weight $w=-k/2$ with a simple pole at the cusp, $
 q\equiv e^{2 i \pi \tau}=0$. This singular behaviour is  associated 
 with the   `unphysical tachyons' which are ubiquitous  in  heterotic vacua,
{\em i.e.} relevant operators in the $(0,2)$-superconformal world-sheet theory which do not respect the level-matching condition. Although $\varPhi(\tau)$ grows exponentially as $\tau\to\I\infty$, the integral \eqref{inthet} is at most power-like divergent, under the condition
that one integrates first on $\tau_1$ and then on $\tau_2$. 
We refer to integrals of the type \eqref{inthet} as `modular integrals with unphysical tachyons'.

For simplicity, we shall restrict ourselves to points in the Narain moduli space 
where the Wilson lines $y_i^a$ vanish, so that the lattice partition function 
factorises as $\varGamma_{(d+k,d)}(g,B,y)=\varGamma_{(d,d)}(g,B)\, \varGamma_{(k,0)}$
where $\varGamma_{(k,0)}$ is a holomorphic modular form of weight $k/2$. At the cost of 
absorbing \footnote{Note that, in general, the limit $y_i^a\rightarrow 0$ is singular due to the appearance of extra massless gauge bosons. However, in the case at hand these massless states do not contribute to the IR behaviour of the  
the modular integral, since we  subtract their infrared divergences. Singularities do however appear at points of symmetry enhancement consistent with the factorisation of the Narain lattice.} 
$\varGamma_{(k,0)}$ into $\varPhi$, we can therefore assume that $k=0$ and $\varPhi$
is a weak holomorphic modular function (i.e. a weak holomorphic modular 
form of weight zero with trivial multiplier system)
\be\label{HolModFormExp}
\varPhi (\tau) = \sum_{n\geq -\kappa} \, a_n\, q^n\ .
\ee
For physics applications, we are only interested in having a simple pole at $q=0$
(i.e. $\kappa=1$), however  our mathematical construction works equally well
for any non-negative integer $\kappa$. 

For $d=0$, the integrand function is holomorphic and the integral \eqref{inthet} is easily 
computed by representing the integrand as a total derivative and using Stokes'  theorem  \cite{Lerche:1987qk,0919.11036}:
\be
\label{Stokesthm}
\int_{\cal F}\de \mu\, \varPhi(\tau) = \frac{\pi}{3}  
\int_{-1/2}^{1/2} \de\tau_1\, G_2 (\tau) \, \varPhi(\tau) = 
\frac{\pi}{3} \left( a_0 - 24 \sum_{-\kappa\leq n < 0} a_n \sigma(-n) \right)\ ,
\ee
where $G_2$ is the holomorphic quasi-modular Eisenstein series of weight two,
and $\sigma(n) = \sigma_1 (n) $ is the sum of divisors of $n$:
\be
G_2 (\tau)= 1-24\sum_{m=1}^{\infty}\frac{m\, q^m}{1-q^m} = 1 - 24 \, \sum_{n=1}^{\infty} \sigma(n)\, q^n\ .
\ee
For $d>0$, however, the integrand function is no longer holomorphic, and cannot be written 
as a total derivative. The standard approach consists instead  in unfolding the lattice sum \cite{Harvey:1995fq,0919.11036}, at the cost of obscuring the T-duality symmetry  
${\rm O}(d,d,\IZ)$ of the Narain partition function. In this section, we shall develop 
techniques for computing integrals of the type \eqref{inthet} while keeping this symmetry manifest. 

Our basic strategy will be to represent  the weak holomorphic modular function 
$\varPhi$ in \eqref{inthet} as a Poincar\'e series, and to unfold {\it it}
rather than the lattice sum\footnote{This is similar in spirit to the Rankin-Selberg method, 
however note that the  Poincar\'e series is no longer an auxiliary function which reduces
to a constant (or a pole with constant residue) at a special value of the parameter, but rather 
represents part of the modular function to be integrated.}. 
While  straightforward in principle, the main difficulty in implementing this idea is the fact that the standard Poincar\'e series representation of weak holomorphic modular functions  
(see e.g. \cite{0695.10021,Manschot:2007ha})
\be
\label{Poincaphi}
\varPhi(\tau) = \tfrac{1}{2} \, a_0 + \tfrac{1}{2} 
\sum_{-\kappa\leq n < 0} a_n
\lim_{K\to\infty} \sum_{|c|\leq K}\quad 
\sum_{|d|<K; (c,d)=1}  \,  e^{2\pi\I n \,\frac{a\tau+b}{c\tau+d}} 
\, \left( 1-e^{\frac{2\pi\I|n|}{c(c\tau+d)}}\right) \ ,
\ee
is not absolutely convergent, so 
the unfolding cannot be justified.

 This problem can be circumvented if the weak holomorphic modular form 
 $\varPhi(\tau)$ can be obtained as a suitable limit of an 
 absolutely convergent non-holomorphic Poincar\'e series
 (or a linear combination thereof),
 \be
\label{defEgen}
E (s,\kappa)\equiv E(\tau_1 , \tau_2 ;s,\kappa) = \tfrac{1}{2} \sum_{(c,d)=1} 
\frac{\tau_2^{s}}{|c\tau+d|^{2s}}\, 
\, e^{-2\pi\I\kappa\, \frac{a\tau+b}{c\tau+d}}\,.
\ee
Here the integers $a$ and $b$ are a solution of $ad-bc=1$. 
The sum over $c,d$ is absolutely convergent for $\Re(s)>1$, and defines an automorphic form of weight $0$.  It may be analytically continued to $s=0$, where the summand 
of \eqref{defEgen} becomes holomorphic, and  the Poincar\'e series \eqref{defEgen}  
formally defines a weak holomorphic modular function which 
behaves as $1/q^\kappa + \cO(1)$ at $q=0$.
On the other hand, for any modular function $F(\tau_1,\tau_2)$,  
the integral over the cut-off fundamental domain 
\be
\label{Rskw}
R_\cT (F,s,\kappa)\equiv \int_{\cF_\cT} \de \mu\, F(\tau_1,\tau_2)\, E (s ,\kappa)
\ee
can be computed for $\Re(s)>1$ by unfolding the sum over $c,d$, and analytically continuing
to $s=0$ as well. Choosing  $F(\tau_1,\tau_2)=\varGamma_{(d,d)} (g,B;\tau )$, and 
assuming that $\varPhi(\tau)$ can be represented as a  linear combination  
\be
\varPhi(\tau) = \lim_{s\to 0} \sum_{-\kappa<n<0} \, a_n\, E (s ,-n)
\ee
of the Poincar\'e series \eqref{Eskw} in the limit $s\to 0$, in principle gives a way to compute the modular integral \eqref{inthet} of interest,
while keeping T-duality manifest. To carry out this program, 
we shall compute the Fourier expansion of  \eqref{defEgen}, and study its limit as $s\to 0$.

Before doing so however, let us first note that
the non-holomorphic Poincar\'e series \eqref{defEgen} is not an eigenmode of the hyperbolic Laplacian
\be
\label{Deltaw}
\Delta= 2 \, \tau_2^2\, \partial_{\bar\tau}\, \partial_\tau\ ,
\ee
but, rather, satisfies \cite{0507.10029}
\be
\label{laplEskw}
\left[\Delta + \tfrac{1}{2}\, s(1-s) \right] \, E(s,\kappa) = 
2\pi\, \kappa \, s \, 
E(s+1,\kappa)\,.
\ee
This equation in principle allows one to determine the analytic properties of 
$E(s,\kappa)$ for $\Re(s)\leq 1$ from the knowledge of the spectrum of 
the Laplacian $\Delta$. 

\subsection{Fourier expansion of the non-holomorphic Eisenstein series\label{sec_fourier}}

The Fourier expansion of $E( s,\kappa)$ can be computed by standard methods \cite{0502.10021,0507.10029,0741.11024,0944.11014,Pribitkin09}.
After extracting the term with $c=0,d=1$, and setting $d=d'+nc$  in the remaining sum,
 a Poisson re-summation over $n$ yields
\be
\label{Eskw}
E( s,\kappa) = \tau_2^{s}\, e^{-2\pi\I \kappa\tau} 
+\sum_{n\in\IZ} \, \tilde E_n (s, \kappa  ) \, e^{2\pi\I n \tau}\,,
\ee
where the $n$-th Fourier coefficient 
\be 
\label{Etilde}
\tilde E_n (s, \kappa  ) = \tau_2^{1-s}\, \sum_{c=1}^{\infty} \frac{S(n,-\kappa;c)}{c^{2s}} \,
A_n (\tau_2 ,c; s,\kappa)
\ee
is expressed through the integral 
\be
\label{defI}
A_n (\tau_2 ,c;s,\kappa) =
\int_{-\infty}^{+\infty} \de t\, 
(t^2+1)^{-s} \ \exp \left[ 2 \pi\I \left( \frac{\kappa}{c^2\tau_2(t+\I)} - n\, \tau_2 \, (t+\I) \right) \right]
\ee
and the Kloosterman sum
\be
S (a,b;c) = \sum_{d\in (\mathbb{Z}/c\mathbb{Z})^*} \exp \left( \frac{2 \pi\I}{c} ( a\, d  + b\, d^{-1})\right)~,
\ee
where $d^{-1}$ stands for the inverse of $d$ modulo $c$ (see Appendix \ref{Kloosterman} for some
relevant properties of the Kloosterman sums). The Kloosterman-Selberg zeta function 
\be
\label{defZklo}
Z (a,b;s) = \sum_{c>0} \frac{S (a,b;c)}{c^{2s}}\ ,
\ee
which is holomorphic for $\Re(s)>1$ and admits a meromorphic continuation to 
all complex values of $s$, will play a central r\^ole in what follows.

The integral  \eqref{defI} can be computed by Taylor expanding $\exp \left( \frac{2\pi\I\kappa}{c^2\tau_2(t+\I)} \right)$, and using the standard formula for the integral
\be
\int_{-\infty}^\infty\, (t+\I)^{-\alpha}\, (t-\I)^{-\beta} \, e^{-2\pi\I u t}\, \de t =
\left\{
\begin{array}{l l}
\frac{2^{2-\alpha-\beta}\pi\, (-\I)^{\alpha-\beta}\, \Gamma(\alpha+\beta-1)}{\Gamma(\alpha)\,\Gamma(\beta)}
& {\rm if}\ u=0\,,
\\
\frac{(2\pi)^{\alpha+\beta}\, (-\I)^{\alpha-\beta}\, u^{\alpha+\beta-1}}{\Gamma(\alpha)\,\Gamma(\beta)}\,e^{-2\pi u}\, \sigma(4\pi u,\alpha,\beta)
& {\rm if}\ u>0\,,
\\
\frac{(2\pi)^{\alpha+\beta}\, (-\I)^{\alpha-\beta}\, (-u)^{\alpha+\beta-1}}{\Gamma(\alpha)\,\Gamma(\beta)}\,e^{2\pi u}\, \sigma(-4\pi u,\beta,\alpha)
& {\rm if}\ u<0\,,
\end{array}
\right.
\ee
valid for $u \in\IR$, $\Re(\alpha+\beta)>1$. Here, the function $\sigma(\eta,\alpha,\beta)$
is related to the Whittaker function $W_{\mu,\nu}(x)$ by
\be
\sigma(\eta,\alpha,\beta) =\varGamma(\beta)\, \eta^{-(\alpha+\beta)/2}\, e^{\eta/2}\, 
W_{\frac{\alpha-\beta}{2},\frac{\alpha+\beta-1}{2}}(\eta)\ .
\ee
In this fashion, we find that the constant term is given by 
\be
\label{Etilde0}
\tilde E_0(s,\kappa) = \sum_{m=0}^\infty
\frac{ 2^{2(1-s)}\, \pi\,   (\pi\kappa)^m\,  \varGamma (2s + m -1)}
{m!\, \varGamma (s + m) \, \varGamma (s)}\, Z(0,-\kappa ; s+m)\,
\tau_2^{1-s-m}\ ,
\ee
while the positive-frequency Fourier coefficients are given by 
\be
\label{Etildep}
\begin{split}
\tilde E_{n>0}(s,\kappa) = \sum_{m=0}^\infty
\frac{(2\pi)^{2(s+m)}\, \, n^{2s+m-1}\, \kappa^m}
{m!\, \varGamma(s)\,\varGamma(s+m)}\,
\sigma\left(4\pi n\tau_2, s+m, s\right)\,
Z(n,-\kappa ; s)\, \tau_2^{s}
\end{split}
\ee
and the negative-frequency Fourier coefficients (besides the first term in \eqref{Eskw})
are given by 
\be
\label{Etildem}
\begin{split}
\tilde E_{n<0}(s,\kappa) = \sum_{m=0}^\infty
\frac{(2\pi)^{2(s+m)}\, (-n)^{2s+m-1}\, \kappa^m}
{m!\, \varGamma(s)\,\varGamma(s+m)}\, 
\, e^{4\pi n \tau_2}\, 
\sigma\left(-4\pi n\tau_2,s, s+m\right)\,
Z(n,-\kappa ; s+m)\, \tau_2^{s}\ .
\end{split}
\ee
Using standard estimates on the Whittaker function and on the 
Kloosterman-Selberg zeta function, one may check that all
these series  are absolutely convergent for $\Re(s)>1$, and therefore
$E( s,\kappa)$ is a meromorphic function of $s$,  analytic 
in the half-plane $\Re(s)>1$. 

It will be useful to rewrite these expressions in a different form. First, using the
fact that the Kloosterman-Selberg zeta function $Z(a,b;s)$ at $a=0$
can be expressed in terms of the Riemann Zeta function via \eqref{Ztozeta},
the constant term  \eqref{Etilde0} may be written as 
\be
\label{Etilde00}
\tilde E_0(s,\kappa) = \sum_{m=0}^\infty
\frac{ 2^{2(1-s)}\, \pi\,   (\pi\kappa)^m\,  \varGamma (2s + m -1)\, \sigma_{1-2s-2m}(\kappa)}
{m!\, \varGamma (s + m) \, \varGamma (s)\, \zeta(2s+2m)}\, 
\tau_2^{1-s-m}\ .
\ee
It may be checked that  \eqref{Etilde00} satisfies the  differential 
equation  \eqref{laplEskw} with $E(s,\kappa)$ replaced by $\tilde E_0(s,\kappa)$.
Second, using the standard
series representation of the Whittaker and Bessel functions
\be
I_\nu (z) = \sum_{m=0}^\infty \frac{(z/2)^{2m+\nu}}{m!\, \varGamma (m+\nu+1)}\ ,\qquad
\sigma(\eta,\alpha,\beta)=\eta^{-\beta}
\, \sum_{p=0}^{\infty} 
\frac{\varGamma(\beta+p)\, \varGamma(1-\alpha+p)}
{p!\, \varGamma(1-\alpha)\, (-\eta)^p} \ ,
\ee
one may carry out the sum over $m$ and obtain, for $n>0$,
\be
\label{Eplus}
\begin{split}
\tilde E_{n} (s,\kappa) =2\pi \,  
\sum_{p= 0}^\infty \frac{\varGamma (s+p)}{\varGamma (s)\, p!} \, 
\frac{2^{-s-p}}{\tau_2^p}\, 
\left( \frac{n}{\kappa} \right)^{\frac{s-p-1}{2}}
\sum_{c>0} \frac{S (n,-\kappa;c)}{c^{s+1+p}}\, I_{s-p-1} \left( \frac{4\pi}{c} \sqrt{\kappa n}\right)\, .
\end{split}
\ee
As an example, we see that the Poincar\'e series $E(s,\kappa)$ can be analytically continued to $s=1/2$, where its Fourier expansion becomes
\be
\label{Ehalf}
E( \tfrac12,\kappa) = \sqrt{\tau_2}\, e^{-2\pi\I \kappa\tau} 
+ \left( 2 \sqrt{\tau_2}\, \sigma_0(\kappa) 
+ \frac{4\pi\kappa}{\zeta(3)} \frac{\sigma_{-2}(\kappa)}{\sqrt{\tau_2}}
+\frac{4\pi^2\kappa^2}{3\zeta(5)} \frac{\sigma_{-4}(\kappa)}{\tau_2^{3/2}}+\dots\right) + \dots~~.
\ee
For $\kappa=1$, this reproduces Eq.(3.45) in \cite{Maloney:2007ud}.

\subsection{Holomorphic limit}

Let us now study the limit $s\to 0$ of the non-holomorphic Poincar\'e 
series $E(s,\kappa)$. At this value, the summand in \eqref{defEgen}
becomes a holomorphic function of $\tau$. Thus, one expects that the analytic continuation $E(0,\kappa)$ will be a holomorphic modular function\footnote{For negative modular weight,
analytic continuation can lead to holomorphic anomalies but, as we shall 
see, these do not occur for $w=0$. We thank J.~Manschot for discussions on this issue.}. In this limit, the constant Fourier coefficient  \eqref{Etilde00} reduces to 
\be
\tilde E_0(0,\kappa) = 12\, \sigma(\kappa)\,.
\ee
Moreover, the negative frequency Fourier coefficients \eqref{Etildem} (besides the 
first term  in \eqref{Eskw})
vanish in this limit,
while the positive frequency Fourier coefficients become
\be
\label{Epluslim}
\tilde E_{n}(\kappa)= 2 \pi\, \, \sqrt{\frac{\kappa}{n}}~ \sum_{c>0} \frac{S (n,-\kappa;c)}{c}\, I_{1} \left( \frac{4\pi}{c} \sqrt{\kappa n}\right)\,.
\ee
Thus, the analytic continuation of $E(s,\kappa)$ at $s=0$ is given by 
\be
\label{Radesum}
E(0,\kappa) = q^{-\kappa} + 12\, \sigma(\kappa) + 
2 \pi\, \sum_{n>0} \, \sqrt{\frac{\kappa}{n}}~ \sum_{c>0} \frac{S (n,-\kappa;c)}{c}\, I_{1} \left( \frac{4\pi}{c} \sqrt{\kappa n}\right)\, q^n\ .
\ee
For $\kappa=1$, this is recognised as the Petersson-Rademacher formula \cite{Petersson,
Rademacher} for the Klein modular invariant $j(\tau)=1/q+196884\,q+\dots$, up to a suitable
additive constant,
\be
\label{Jexp}	
E (0,1) = j(\tau) + 12\,.
\ee
For $\kappa>1$, the r.h.s. of \eqref{Radesum} is obtained from its value at $\kappa=1$ 
by acting with the 
the Hecke operator
\be
\label{defHecke}
(T_\kappa \cdot \varPhi)(\tau)=\sum\limits_{\substack{a,d>0 \\ ad=\kappa}}~~ \sum_{b\,\textrm{mod}\, d} 
\varPhi\left( \frac{a\tau+b}{d}\right)\ ,
\ee
which acts in the space of weak holomorphic modular functions and maps 
$\varPhi=1/q+\cO(q)$ to $T_\kappa \cdot \varPhi=1/q^\kappa+\cO(q)$. 
This can  be checked  using  Selberg's identity \eqref{selbergidentity} 
for the Kloosterman sum and the fact that $T_\kappa$ acts on the Fourier coefficients as 
\be
(T_\kappa \cdot \tilde E)_n = \kappa \,\sum_{d|(n,\kappa)} d^{-1}\,\tilde E_{n \kappa/d^2} \,.
\label{heckeop}
\ee
Thus, 
the analytic continuation of $E(s,\kappa)$ at $s=0$ is, in general, the weak
holomorphic modular function
\be
\label{Jexpk}
E (0,\kappa) =  T_\kappa \cdot j(\tau) + 12\, \sigma(\kappa)\,.
\ee
Indeed, a numerical evaluation of \eqref{Radesum} 
reproduces the known coefficients of the $j$-function and its images under $T_\kappa$.

\subsection{Shifted constrained Epstein zeta series}

In this subsection we shall evaluate the modular integral \eqref{Rskw} for an arbitrary modular function $F(\tau_1,\tau_2)$ of moderate growth. In the interest
of simplicity, and with a view towards our main goal \eqref{inthet}, we assume that 
\be
\label{assumF}
F_0(\tau_2)=\tau_2^{d/2}\,,
\ee
and we require that there exists an $a>0$ such that
\be
\label{assumFk}
F_{\kappa}(\tau_2) \, e^{2\pi\kappa\tau_2} \sim e^{-a\tau_2}
\ee
as $\tau_2 \to \infty$. Here $F_{\kappa}(\tau_2)
=\int_{-1/2}^{1/2} F(\tau_1,\tau_2)\, e^{-2\pi\I \kappa\tau_1}\, \de\tau_1$ 
is the $\kappa$-th Fourier coefficient of $F$. Notice that these conditions are satisfied by the
lattice partition function $\varGamma_{(d,d)}$ away from points of extended 
gauge symmetry. 

By the same reasoning as in eqs. \eqref{sumsac}-\eqref{Zagier0}, for large enough $\Re(s)$ 
we can unfold the sum over $(c,d)$ in the non-holomorphic Poincar\'e series \eqref{defEgen}, arriving at 
 \be \label{Prescription}
R_\cT (F,s,\kappa ) = 
 \int_0^\cT\, \de\tau_2\, \tau_2^{s-2} e^{2\pi\kappa\tau_2}\,   F_{\kappa}(\tau_2)  
   - \int_{\cF-\cF_{\cT}} \de\mu \, F (\tau_1 , \tau_2)\,
\left( E(s,\kappa ) -  \tau_2^{s} \, e^{- 2\pi \I \kappa\tau} 
\right)  \, .
\ee
By our assumption \eqref{assumF}, the term on the first line is finite as $\cT\to\infty$.
The term on the second line can be decomposed into  
\be
- \int_{\cF-\cF_{\cT}} \de\mu \, \left( F (\tau_1 , \tau_2) - \tau_2^{d/2}\right) \,
\left( E(s,\kappa) -  \tau_2^{s} \, e^{- 2\pi \I \kappa\tau} \right)
- \sum_{m=0}^{\infty} \, e_m(s,\kappa)\,  \int_{\cT}^{\infty} \de\tau_2\,   \tau_2^{-1-s-m+\frac{d}{2}} 
 \, ,
\ee
where 
\be
\label{emskw}
 e_m(s,\kappa)=
2^{2(1-s)}\, \pi\, 
\frac{  (\pi\kappa)^m\, \sigma_{1-2s-2m}(\kappa)\, \varGamma (2s + m -1)}{m!\, 
\varGamma (s + m) \, \varGamma (s)\, \zeta(2s+2m)}~.
\ee
The first term in this expression
is exponentially suppressed in the limit $\cT\to\infty$, thus, up to exponentially suppressed
terms, the modular integral \eqref{Rskw} is given by
\be
\label{RTskw}
R_\cT (F,s,\kappa) = 
\int_0^{\infty}\, \de\tau_2\, \tau_2^{s-2} e^{2\pi\kappa\tau_2}\,   F_{\kappa}(\tau_2) 
+\sum_{m=0}^{\infty} \frac{\cT^{\frac{d}{2}-s-m}\, e_m (s,\kappa)}{\frac{d}{2}-s-m} +\ldots ~~.
\ee
For $\Re(s)>d/2$, the second term in \eqref{RTskw} is suppressed as 
$\cT\to\infty$, and the cut-off integral \eqref{Rskw} converges to the first term in 
\eqref{RTskw} as the cut-off is removed. We define the renormalised integral as the
analytic continuation of this result to all $s$, namely by the Mellin transform
\be
\label{defrenormIk}
{\rm R.N.} \, \int_{\mathcal F} F(\tau_1,\tau_2)\, E(s,\kappa) \,d\mu \equiv 
\int_0^{\infty}\, \de\tau_2\, \tau_2^{s-2} e^{2\pi\kappa\tau_2}\,   F_{\kappa}(\tau_2) \ .
\ee
As in the case considered in Section 2, the analytic structure of \eqref{defrenormIk} can be read 
off from the growth of $e^{2\pi\kappa\tau_2}\,   F_{\kappa}(\tau_2)$ near 
$\tau_2\to\infty$. Unlike  the case studied in Section 2, the integral \eqref{defrenormIk} is not expected to obey a simple
functional equation under $s\to 1-s$. 

Let us now specialise to the case $F (\tau_1 , \tau_2) =\varGamma_{(d,d)}$. The 
$\kappa$-th Fourier coefficient is given by 
\be
F_\kappa(\tau) = \tau_2^{d/2}\,
\sum_{p_L, p_R}\, e^{-\pi\tau_2 \left( p_L^2 + p_R^2\right)} \delta \big( p_L^2 - p_R^2-2\kappa\big)\ ,
\ee
where the sum is restricted to lattice vectors satisfying the usual level-matching constraint $p_L^2 - p_R^2=2\kappa$.
For such vectors,  $p_L^2 + p_R^2=2 (p_R^2+\kappa)$,  and the condition 
\eqref{assumFk} is therefore obeyed away from loci on the symmetric space 
${\rm O}(d,d,\IR)/{\rm O} (d)\times {\rm O} (d)$ where $p_R^2$ vanishes for some lattice
vector satisfying the constraint above. Physically, these loci correspond to points
of enhanced gauge symmetry, where additional massless states occur, leading
to new infrared divergences. We shall always work away from such points.

Under this assumption, the Mellin transform \eqref{defrenormIk} can be computed explicitly
by integrating term by term, as in \eqref{REps}. The result is a `shifted constrained 
Epstein zeta series'
\be
\label{intEskwd}
{\rm R.N.}
\int_{\cF}\de\mu\, \varGamma_{(d,d)}(g,B) \, E( s,\kappa) 
= \frac{\varGamma (s+\frac{d}{2}-1)}{\pi^{s+\frac{d}{2}-1}} \, \mathcal{E}^{d}_V (g,B;
s+\tfrac{1}{2}d-1,\kappa) \,,
\ee
where
\be
\label{defsEpstein}
\mathcal{E}^{d}_V (g,B;s,\kappa) \equiv 
{\sum_{p_L,p_R}} \frac{\delta \big( p_L^2 - p_R^2-2\kappa\big)}
{(p_L^2+p_R^2-2\,\kappa)^{s}} \,.
\ee
This series is absolutely convergent when $\Re(s)>d$.
Unlike the `unshifted' Epstein zeta series  \eqref{defceps},  which we henceforth denote
by $\mathcal{E}^{d}_V (g,B;s,0)$, the shifted series \eqref{defsEpstein}
diverges
as $(2 p_R^2)^{-s}$
at the points of enhanced gauge symmetry where the norm $p_R^2$ of 
some lattice vector satisfying the constraint vanishes. 
Moreover, using the differential
equation \eqref{DiffEqEisensteina}, satisfied by the lattice partition function, together with the differential equation \eqref{laplEskw} satisfied by the 
non-holomorphic Poincar\'e series, 
we find that the shifted constrained Epstein zeta series satisfies 
\be
\label{LapsEpstein}
\left[ \Delta_{SO(d,d)} -s(s-1) + \tfrac14 \,d(d-2) \right]\, 
\mathcal{E}^{d}_V (g,B;s+\tfrac{1}{2}\, d-1,\kappa) = 2\pi\, \kappa\, s \, 
\mathcal{E}^{d}_V (g,B;s+\tfrac{1}{2}\, d ,\kappa) ~.
\ee
In the limit $s\to 0$,  the r.h.s. vanishes and therefore 
$\mathcal{E}^{d}_V (s+\frac{d}{2}-1,\kappa)$
is an eigenmode of the Laplacian on ${\rm O} (d,d,\IR)/ {\rm O} (d)\times {\rm O}(d)$
with eigenvalue $d(2-d)/4$. This argument assumes that 
$\mathcal{E}^{d}_V (s+\frac{d}{2},\kappa)$ is
finite at $s=0$, otherwise the r.h.s. of \eqref{LapsEpstein} may be non-vanishing.

\subsection{Holomorphic limit of the integral\label{homlin}}

Let us now investigate the limit of the integral \eqref{intEskwd} at $s=0$.
As discussed below eq.\,\eqref{defEgen}, the non-holomorphic Poincar\'e series becomes
holomorphic in this limit,  and one may hope to recover a
modular integral of the form \eqref{inthet}. Of course, infrared divergences require
special care. 

Returning to \eqref{RTskw} at finite infrared cut-off $\cT$, we see that all power-like
terms vanish in the limit $s\to 0$, appart from those corresponding to $m=1$ (whose
coefficient goes to a constant $e_1(s,\kappa)=12 \sigma(\kappa)$, c.f. \eqref{Jexp})
and to $m=d/2$ (when $d$ is even), whose coefficient $e_m$ vanishes linearly 
in $s$ but which are divided by a vanishing number. Considering first the
case where $d$ is odd and greater than $3$, we thus have
\be
\label{limint1}
\lim_{s\to 0} \, \left(  \int_{\cF_\cT}\, \de\mu\, E (s,\kappa) \, \varGamma_{(d,d)} \right)
= \frac{\varGamma (\frac{d}{2}-1)}{\pi^{\frac{d}{2}-1}} \, \mathcal{E}^{d}_V (g,B;\tfrac{1}{2}\, d-1,\kappa) + 
 12 \sigma(\kappa)\, \frac{\cT^{\frac{d}{2}-1}}{\frac{d}{2}-1}~.
\ee 
On the other hand, in the limit $s\to 0$, $E (s,\kappa)$ reduces to 
the holomorphic modular forms $T_\kappa \cdot j(\tau) + 12\, \sigma(\kappa)$, cf. \eqref{Jexpk}.
The constant term in this expression is responsible for the divergent term in \eqref{limint1}. After subtracting this divergence, we conclude that 
\be
{\rm R.N.} \int_{\cF}\, \de\mu\, \left[ T_\kappa \cdot j(\tau) + 12\, \sigma(\kappa) \right] \, \, \varGamma_{(d,d)} 
= \frac{\varGamma (\frac{d}{2}-1)}{\pi^{d/2-1}} \, \mathcal{E}^{d}_V (g,B;\tfrac{1}{2}\, d-1,\kappa) \ ,
\ee
or, using \eqref{REps},
\be
{\rm R.N.} \int_{\cF}\, \de\mu\, T_\kappa \cdot j(\tau) \, \, \varGamma_{(d,d)} 
= \frac{\varGamma (\tfrac{d}{2}-1)}{\pi^{d/2-1}} 
\left[  \mathcal{E}^{d}_V (g,B;\tfrac{1}{2}\, d -1,\kappa) \,
-12\, \sigma(\kappa)\, 
\mathcal{E}^d_V (g,B;\tfrac{1}{2}\, d-1,0) \right]~.
\ee
If $d$ is even and $d\geq 4$, there is an additional constant term on the r.h.s. of \eqref{limint1},
coming from $m=d/2$:
\be
\lim_{s\to 0} \, \left(  \dots \right)
= \frac{\varGamma (\frac{d}{2}-1)}{\pi^{\frac{d}{2}-1}} \, \mathcal{E}^{d}_V (g,B;\tfrac{1}{2}\, d-1,\kappa) + 
 12 \, \sigma(\kappa)\, \frac{\cT^{\frac{d}{2}-1}}{\frac{d}{2}-1}
 + \frac{8\pi(\pi \kappa)^{d/2}\, \sigma_{1-d}(\kappa)}{(d/2)!\, (d-2)\, \zeta(d)}~.
\ee
After subtracting the divergent term, we conclude that for $d\geq 4$ even,
\be
\begin{split}
{\rm R.N.} \int_{\cF}\, \de\mu\, T_\kappa \cdot j(\tau) \, \, \varGamma_{(d,d)} 
= & \frac{\varGamma (\tfrac{d}{2}-1)}{\pi^{d/2-1}} 
\left[  \mathcal{E}^{d}_V (g,B;\tfrac{1}{2}\, d-1,\kappa) \,
-12\, \sigma(\kappa)\, 
\mathcal{E}^d_V (g,B;\tfrac{1}{2}\, d-1,0) \right]\\
&+ \frac{8\pi(\pi \kappa)^{d/2}\, \sigma_{1-d}(\kappa)}{(d/2)!\, (d-2)\, \zeta(d)} \,.
\end{split}
\ee
In the remainder of this subsection we deal with the low dimensional cases. 

For $d=0$, both $m=0$ and $m=1$ contribute, leading to 
\be
\int_{\mathcal{F}_\cT} \de\mu\, \left[ T_\kappa \cdot  j(\tau) + 12 \sigma(\kappa) \right] = 
-12\sigma(\kappa)\cT^{-1} - 4\pi \sigma(\kappa)+ \dots~~.
\ee
Using $\int_{\mathcal{F}_\cT}{d\mu}=\frac{\pi}{3}-\frac{1}{\cT}$, we arrive at 
\be
\label{jint}
\int_{\cal F_\cT} \de\mu\, T_\kappa \cdot j(\tau ) = - 8 \pi\,\sigma(\kappa)\, ,
\ee
which agrees with the result of  Stokes' theorem \eqref{Stokesthm}. Notice that this result is independent of the cutoff $\cT$, since the only contribution to the modular integral 
comes from the region $\tau_2\leq 1$ of ${\cal F}$.
 
For $d=1$, only $m=1$ contributes, leading to 
\be
\int_{\mathcal{F}_\cT} \de\mu\, \left[ T_\kappa \cdot  j(\tau) + 12 \sigma(\kappa) \right] \, 
\varGamma_{(1,1)} (R) = -24\, \sigma(\kappa)\cT^{-1/2} - 4\pi \, \sum_{\substack{m,n>0\\ mn=\kappa}}\, |m/R-nR|
\ee
and therefore
\be
\label{intj11}
\int_{\mathcal{F}} \de\mu\, T_\kappa \cdot  j(\tau)  \, \varGamma_{(1,1)} (R)
= 
 - 4\pi \, \sum_{\substack{m,n>0\\ mn=\kappa}}\, |m/R-nR| - 4\pi\,  \sigma(\kappa) (R+1/R)\ .
\ee
For $R>\sqrt{\kappa}$, Eq. \eqref{intj11} reduces to $-8\pi \sigma(\kappa) R$, which would be the
result of naively applying the unfolding trick to the partition sum $\varGamma_{(1,1)} (R)$, as in \cite{Bachas:1997mc}. Similarly, for $R<1/\sqrt{\kappa}$, Eq. \eqref{intj11} reduces to $-8\pi \sigma(\kappa)/ R$, as required by T-duality. For $1/\sqrt{\kappa}<R<\sqrt{\kappa}$, however, the naive unfolding of $\varGamma_{(1,1)} (R)$ fails, while our method still applies. 
We note that the Eq. \eqref{intj11} for $\kappa>1$ can be derived from the $\kappa=1$ result 
by observing that the lattice partition function satisfies\footnote{This identity is easily established for $\kappa$ a prime number, and can be extended to the general case using the Hecke algebra $T_\kappa\, T_{\kappa'} =  \sum_{d|(\kappa,\kappa')} d\, T_{\kappa\kappa'/d^2}$ (note the 
non-standard normalization of the Hecke operators in \eqref{defHecke}). }
\be
\label{Tg11}
T_\kappa\cdot \varGamma_{(1,1)} (R;\tau) =
 \sqrt{\kappa} \, \sum_{\substack{m,n>0\\ mn=\kappa}}\,  \varGamma_{(1,1)} (R\sqrt{m/n};\tau) \ ,
\ee
and invoking the Hermiticity of the Hecke operator $T_\kappa$ with respect to the Petersson product. Eq. \eqref{Tg11}  may be viewed as a $p$-adic analogue (for $\kappa=p$ prime) of \eqref{DiffEqEisensteina}.

Finally, we consider the most complicated case $d=2$. In this case the 
coefficient $e_1$ behaves as 
\be
e_1 = \alpha + \beta s +\cO(s^2)\ ,
\ee
with
\be
\alpha= 12 \sigma(\kappa) \ ,\qquad
\beta = -24\, \sigma(\kappa) \left[ \gamma -12 \log A + \log(4\pi)\right] -24\,\kappa\sigma'_{-1}(\kappa) 
\ ,
\ee
where $A$ is the Glaisher constant. After integrating over $\tau_2$, we arrive at 
\be
\int_{\cF_\cT}\, \de\mu\, E (s,\kappa) \, \varGamma_{(2,2)}  
= 
 \frac{\varGamma (s)}{\pi^{s}} \, \mathcal{E}^{2}_V (g,B;s,\kappa) 
-\frac{\alpha}{s}\,
+\alpha\log\cT -\beta+ \cO(s)\ .
\ee
The pole and logarithmic divergences on the r.h.s. originate from the constant term
$12\, \sigma(\kappa)\, \tau_2^s$ in the limit $s\to 0$ of $E (s,\kappa)$. After subtracting these
terms, we conclude that 
\be
\label{intj2k}
\int_{\mathcal{F}} \de\mu\, T_\kappa \cdot  j(\tau)  \, 
\varGamma_{(2,2)} 
= \lim_{s\to 0}  \left[\, \frac{\varGamma (s)}{\pi^{s}} \, \mathcal{E}^{2}_V (g,B;s,\kappa)  \right]
- \beta -12\, \sigma(\kappa)\, I_2\,.
\ee
In particular, we conclude the Epstein zeta series  $\mathcal{E}^{2}_V (g,B;s,\kappa)$
has a zero at $s=0$.
 
On the other hand, the integral \eqref{intj2k} may be determined using harmonicity
and the singularity structure near points of enhanced symmetry. E.g. 
for $\kappa=1$, states with momenta $m_1=n_1=0$, $m_2=-n_2=\pm 1$
become massless\footnote{Note that 4 additional massless states appear at 
the special values $T=U=i$ and $T=U=\rho$, where $\rho=e^{i\pi/3}$.}  
at $T=U$, and induce a singularity of the form 
\begin{align}
    -4\log|T-U|~.
\end{align}
More generally, massless states arise whenever $j(T)=j(U)$, 
as required by invariance under the T-duality group ${\rm SL} (2,\mathbb{Z})_T\times {\rm SL} (2,\mathbb{Z})_U\ltimes\mathbb{Z}_2$.  Combining this singularity structure with  
harmonicity, one can uniquely determine
 \be
 \label{HMBor}
\int_{\mathcal{F}} \de\mu~  j(\tau)  \, 
\varGamma_{(2,2)} (T,U) 
= -\log\, |j(T)-j(U)|^4 + {\rm const}~.
\ee
This result was confirmed in \cite{Harvey:1995fq} by using the unfolding trick and 
of Borcherds' product formula for $j(T)-j(U)$. 

Similarly to the one-dimensional case, the integral \eqref{intj2k}\ can be evaluated for $\kappa>1$ by observing that the lattice partition function satisfies the $p$-adic analogue of \eqref{Diff2s}
\begin{equation}
T_\kappa^{(\tau)} \, \varGamma_{(2,2)} =
T_\kappa^{(T)} \, \varGamma_{(2,2)}  =
T_\kappa^{(U)} \, \varGamma_{(2,2)} \,,
\end{equation}
where $T_\kappa^{(T,U)}$ are the analogues of the Hecke
operator  \eqref{defHecke}, now acting on the moduli $T$ and $U$.
Invoking the Hermiticity of $T_\kappa$ with respect to the Petersson product, one
finds that \eqref{HMBor} generalizes into 
 \be
 \label{HMBor2}
\int_{\mathcal{F}} \de\mu\, T_\kappa \cdot  j(\tau)  \, 
\varGamma_{(2,2)} (T,U) 
= -\tfrac12 (T_\kappa^{(T)} + T_\kappa^{(U)} ) \log\, |j(T)-j(U)|^4 + {\rm const}\,.
\ee
This result is consistent with the fact that additional massless states arise
at $T=\kappa\, U$, together with the images of this locus under T-duality.

Comparing \eqref{HMBor2} with \eqref{intj2k}, we arrive 
at interesting identities between the constrained Epstein zeta series $\mathcal{E}^{2}_V (g,B;s,
\kappa)$ at $s=0$ and the $j$ function. It would be interesting to compute the 
constrained Epstein zeta series directly, perhaps along the lines of Section \ref{sec_lowdim}, and
derive these identities independently. 

More generally, it will be interesting to extend our method to modular integrals of
non-symmetric lattice partition functions $\varGamma_{d+k,d}$. While the generalisation
of the non-holomorphic Poincar\'e series \eqref{defEgen} to negative weight is obvious, 
its analytic continuation to the relevant value of $s$ where the summand becomes 
holomorphic is subtle, and involves an interesting interplay of holomorphic and modular 
anomalies \cite{Pribitkin09}. We hope to discuss this problem in future work.

\subsection*{Acknowledgements}

We are grateful to S. Stieberger for asking a question which prompted this project, and to S. Hohenegger, 
E. Kiritsis, C. Kounnas, J. Manschot, S. Murthy and K. Siampos for interesting comments and discussions. I.F. would like to thank the University of Torino for hospitality and
 C.A. and I.F. would like to thank the TH Unit at CERN for hospitality during the final stages of the project. This work was partially supported by the European ERC Advanced Grant no. 
226455 ``Supersymmetry, Quantum Gravity and Gauge Fields'' (SUPERFIELDS), the ERC Advanced Grant 226371 ``MassTeV'', by the CNRS PICS Nos. 3747 and 4172, and by the Italian MIUR-PRIN 
contract 2009KHZKRX-007 ``Symmetries of the Universe and of the Fundamental Interactions''.


\appendix

\section{Properties of Kloosterman sums}\label{Kloosterman}

Kloosterman sums  play an central r\^ole in Number Theory. The classical Kloosterman sums for the modular group $\varGamma = {\rm SL} (2,\mathbb{Z})$ are defined by
\be
S (a,b;c) = \sum_{d\in (\mathbb{Z}/c\mathbb{Z})^*} \exp\left[ \frac{2\pi \I}{c} (a \, d+ b\, d^{-1})\right]\,,
\ee
where $a$, $b$ and $c$ are integers, and $d^{-1}$ is the inverse of $d$ mod $c$, and enter in the explicit expression of the Fourier coefficients of modular forms. $S (a,b;c)$ is clearly symmetric under the exchange of $a$ and $b$. Less evidently, it satisfies the Selberg identity
\be 
\label{selbergidentity}
S (a,b;c) = \sum_{d | {\rm gcd} (a,b,c)} d\, S ( ab/d^2,1;c/d)\,.
\ee
In the special case $a\neq 0$, $b=0$, the Kloosterman sum reduces to the Ramanujan sum
\be
S(a,0;c) = S(0,a;c) = \sum_{d\in (\mathbb{Z}/c\mathbb{Z})^*} \exp \left( \frac{2\pi \I}{c} a\, d \right) = \sum_{d | {\rm gcd} (c,a)}d\,  \mu (c/d)\,,
\ee
with $\mu (n)$ the M\"obius function. For $a=b=0$, $S (a,b;c)$ reduces instead to the 
Euler totient function $\phi(c)$. 

We now turn to the Kloosterman-Selberg zeta function \eqref{defZklo}. Using
the trivial bound $|S(a,b;c)|<c$, one sees immediately 
that the sum over $c$ converges absolutely when $\Re(s)>1$. The Weil bound 
$|S(a,b;c)|<2^{\nu(c)}\, \sqrt{c \, gcd(a,b,c)}$, where $\nu(n)$ is the number of divisors of $n$, shows that $Z(a,b;s)$ is in fact analytic when $\Re(s)>3/4$.
When one or both of the
arguments vanish, it can be expressed in terms of   the Riemann zeta function via
\be
\label{Ztozeta}
Z(0,0;s) = \frac{\zeta(2s-1)}{\zeta(2s)}\,,\qquad 
Z (0,\pm \kappa ;s ) = \frac{\sigma_{1-2s} (\kappa )}{\zeta (2s)}
\qquad (\kappa\neq 0)~,
\ee
where $\sigma_s (n)$ is the divisor function
\be
\label{defsigapp}
\sigma_s(n) = \sum_{d|n} d^s\,.
\ee 
The usual notation $\sigma(n)\equiv \sigma_1(n)$ is used throughout the paper.



\providecommand{\href}[2]{#2}\begingroup\raggedright\endgroup

\end{document}